\documentclass[aps, preprint,amsmath,amssymb,eshowkeys,nofootinbib]{revtex4-1}

\usepackage{verbatim,graphics,graphicx,color,slashed,textcomp,bbm,mathdots,multirow,array}
\usepackage{dcolumn}
\usepackage{bm}
\usepackage{revsymb4-2}
\usepackage{autobreak}

\graphicspath{{figures/}}
\usepackage[colorlinks=true,linkcolor=red,urlcolor=blue,citecolor=blue]{hyperref}

\begin{document}


\title{Velocity Distribution of Dark Matter in Spikes around Schwarzschild Black Holes and Effects on Gravitational Waves from EMRIs}


\author{Zi-Chang Zhang$^{a}$} 

\author{Yong Tang$^{a,b,c}$}
\affiliation{\begin{footnotesize}
		${}^a$School of Astronomy and Space Science, University of Chinese Academy of Sciences (UCAS), Beijing 100049, China\\
		${}^b$School of Fundamental Physics and Mathematical Sciences, \\
		Hangzhou Institute for Advanced Study, UCAS, Hangzhou 310024, China \\
        ${}^c$International Center for Theoretical Physics Asia-Pacific, Beijing/Hangzhou, China 
		\end{footnotesize}}

\date{\today}

\begin{abstract}
Dark matter (DM) constitutes the predominant portion of matter in our universe. Despite compelling evidence, the precise characteristics of DM remain elusive. Among the leading DM candidates are weakly-interacting massive particles, which may clump into steep concentrations around the central black holes of galaxies. However, DM profiles of the resulting dense spikes remain uncertain. Here we employ the relativistic dynamics in Schwarzschild geometry and first evaluate the velocity distributions of DM within such spikes. Through variations in black hole masses and dark halo parameters, we identify universal features in DM profiles and fit them with Gaussian distributions. Additionally, we illustrate with the impact of dynamical friction on gravitational waves generated by extreme-mass-ratio inspirals (EMRIs) within DM spikes, taking into account the velocity distribution of DM in the relativistic regime. Our findings demonstrate the phase shifts in the time-domain waveform, potentially providing useful insights for probing DM in galactic centers by gravitational-wave experiments.


\end{abstract}


\maketitle

\section{Introduction} \label{sec:intro}





Various observational evidence suggests that DM constitutes a substantial component of the universe, including galactic rotational curves, colliding bullet cluster, gravitational lensing, large-scale structure, and anisotropy in cosmic microwave background.
However, the exact nature of DM is still unknown.
One leading paradiam suggests that DM might be weakly-interacting massive particles (WIMPs), which have been searched on particle colliders, directly with scattering with atoms, and indirectly by annihilation relics in astrophysical environments \cite{BERTONE2005279,doi:10.1142/S0217751X13300408,Schumann_2019}. And the latter two methods depend on DM density profile.

To describe the density distribution of DM in galaxies, different models have been proposed, such as power-law profile, Navarro-Frenk-White (NFW) profile \cite{1996ApJ.462.563N,Navarro_1997} and Hernquist profile \cite{1990ApJ.356.359H}. 
These models describe the mass density of galactic DM well, while still leaving large uncertainty in galactic centers which are usually accompanied by massive black holes. In \cite{PhysRevLett.83.1719} (hereafter referred to as GS) it is found that DM will be accreted and redistributed in the vicinity of a black hole, forming a high-density spike that might be observable through gamma rays from DM annihilation. 
Moreover, fully relativistic spikes around Schwarzschild black holes \cite{PhysRevD.88.063522} and Kerr black holes \cite{PhysRevD.96.083014} were investigated (hereafter referred to as SFW and FRW, respectively), confirming the presence of spike but with different profiles.

To probe the density profile of DM spike, gravitational wave (GW) might serve as a feasible new way. The inaugural discovery of GW \cite{PhysRevLett.116.061102} 
has significantly invigorated studies in black hole physics and GWs from supermassive black holes (ranging from $10^4 M_\odot$ to $10^9 M_\odot$). Concurrently, these topics form the cornerstone of upcoming space-based GW observatories like LISA \cite{amaroseoane2017laser, baker2019laser}, Taiji \cite{10.1093/nsr/nwx116,Ruan:2020smc} and TianQin \cite{Luo_2016}. 
The seminal work by \cite{1943ApJ.97.255C} on the general dynamical friction (DF) established that the frictional force is proportional to the density of the surrounding environment, which has been utilized in recent discussions on using GWs to probe DM near black holes \cite{PhysRevLett.110.221101,10.21468/SciPostPhysCore.3.2.007,PhysRevD.102.083006,PhysRevD.105.043009,article,zhang2024detecting} and extended to wider context~\cite{Traykova:2021dua, Traykova:2023qyv, Bamber:2022pbs, Boudon:2022dxi, Boudon:2023qbu, Liang:2022gdk, Boey:2024dks}. 
As a compact object orbits around the central black hole, it is subjected to DF from the surrounding DM, which changes the orbital trajectory and furthermore affects the GWs emitted by such a binary \cite{Macedo_2013}. 
{For searching DM in galactic center, notable observational targets are intermediate-mass-ratio inspirals (IMRIs) and extreme-mass-ratio inspirals (EMRIs),
the binaries consisting of a supermassive black hole and a compact object with a much smaller mass. 
The typical GW frequencies of IMRIs and EMRIs lay in the ranges of LISA, Taiji, TianQin. 
Therefore, detecting IMRIs and EMRIs is one of the main missions of space-based GW detectors. 
In the meanwhile, the supermassive black holes in IMRIs and EMRIs tend to be located at the center of galaxies, which reveals the environment in galactic center and offers the possibility of detecting DM spikes.}
It was demonstrated that DM minispikes can elevate the merger rate of IMRIs \cite{Yue_2019}, potentially affecting the detectable event rate by LISA. 
Additionally, the impact of DM spikes on stellar orbits in the Galactic center has been explored \cite{shen2023exploring}, and in the future higher-resolution S-star observations will provide more accurate estimations of the parameters of the DM spikes.
Although it was mentioned that galactic merging \cite{PhysRevLett.88.191301} 
and stellar motion \cite{PhysRevLett.92.201304,PhysRevD.72.103502} might kick out and disrupt the DM profiles in galaxies, 
in scenarios where massive black holes have avoided major mergers, DM minihalos can coexist nearby \cite{PhysRevLett.95.011301}. 

Considerable efforts had been devoted to the study of orbital dynamics and GW emission based on 
power-law profile or fitting adiabatic growth formula for the density of DM generating DF \cite{PhysRevD.97.064003,PhysRevD.100.043013,PhysRevD.102.103022,Tang_2021, PhysRevD.106.044027}. 
However, these discussions ignored the velocities of the DM particles. 
In fact, the effect of velocity distribution plays an important role because Chandrasekhar's formula suggests that only particles below the orbital velocity of compact bodies are contributing DF. In this regard, \cite{2022SCPMA.6500412L} and \cite{PhysRevD.105.063029} discussed the effect of particle velocity distribution on orbital circularity or ellipticity, and showed that high-speed particles do not participate in gravity drag, which inevitably weakens DF. 
Besides, \cite{karydas2024sharpening} introduced the circularizing effects of accretion of DM on secondary objects eccentricity evolution 
and the velocity distribution of DM was considered in numerical simulations \cite{kavanagh2024sharpening}.
However, relativistic effects near black holes were rarely considered, such as the relativistic density and velocity distribution after adiabatic growth. Because near the horizon of a black hole or when the binary is close to merge, we expect the system is generically relativistic and it will be more consistent to take such an effect into consideration. 

{In this paper, differently from the Newtonian analysis of GS, we calculate the velocity distributions of relativistic DM in spikes formed by the adiabatic growth of Schwarzschild black holes from initial Hernquist dark halos or power-law dark halos.}
We employ the geodesic equations in the numerical calculations to account for the velocities of DM particles, and categorize them accordingly, leading to the identification of spikes formed by DM particles up to different maximum velocities. {We also evaluate velocity dispersion and fit with the Gauss distribution.} Then, incorporating velocity into the relativistic DF formula, we illustrate with the GWs of EMRIs up to 3PN (post-Newtonian) eccentric gravitational radiation and demonstrate the phase shifts of waveforms in time domain. 

This paper is structured as follows. In Section \ref{sec:DM}, we elaborate on how the adiabatic growth of the black hole modifies the original dark halo profile. This involves considering the conditions for black hole capture and evaluating the velocity distribution based on particle geodesic velocity.
In Section \ref{sec:GW}, we shed light on the velocity distribution and the energy-momentum loss resulting from GW emission and DF from DM spikes. Then we calculate the GW signals by integrating the geodesic equations of the compact object to derive the orbital trajectories. Finally in Section \ref{sec:RE} we show the numerical results, and summarize in Section \ref{sec:CON}.

\section{DARK MATTER SPIKE} \label{sec:DM}

We first outline the relevant theoretical formalism and conventions~\cite{PhysRevD.88.063522}, including the relativistic dynamics and kinematics in the Schwarzschild metric. After obtaining the analytic formulas, we shall solve these equations numerically in later sections.  

\subsection{DM growth around Schwarzschild black hole} \label{subsec:growth}
For a system of DM particles with given relativistic phase space distribution $f^{(4)}(x,p)$, the mass current four-vector can be expressed as 
\begin{equation}
J^\mu(x)=\int f^{(4)}(x,p)u^\mu \sqrt{-g} d^4p ,
\end{equation}
where $p$ is the four-momentum, $u^\mu = {p^\mu}{/}{\mu}$ is the four-velocity, 
$\mu$ is rest mass,  
$g$ is the determinant of the metric, and $d^4p$ is the four-momentum volume element. 
The distribution function $f^{(4)}(x,p)$ of DM is normalized, $\int f^{(4)}(x,p)\sqrt{-g} d^4p = 1$.

Using the relations $J^\mu=\rho u^\mu$ and $u_\mu u^\mu=-1$, we can obtain the density in a local free-falling frame $\rho = \sqrt{-J_\mu J^\mu}$. Therefore, the estimation of the components of $J^\mu$ is the main task. 
In general we can get the distribution function by Eddington formula 
\cite{1916MNRAS.76.572E} with a given density. However, it is not useful here because the density is unknown. 
In addition, the integral boundary of four-momentum volume element is undetermined, which results in the difficulty to integrate directly. 
Fortunately, we can overcome these by converting the integral element $d^4p$ into the phase space of invariant constants of motion and leveraging the adiabatic invariance of the distribution function. Then using numerical integration we can obtain the components of $J^\mu$. 


The line element in Schwarzschild metric for a black hole of mass $M$ (with the speed of light $c=1$) can be written as
\begin{equation}
ds^2 = -(1-\frac{2GM}{r})dt^2 + \frac{r}{r-2GM}dr^2 + r^2d\theta^2 +r^2\sin^2{\theta}d\phi^2 .
\end{equation}
Due to spherical symmetry, four constants of motion admitted by timelike geodesics are
\begin{equation}\label{eqs3}
\begin{aligned}
 \varepsilon &\equiv -u_t = -g_{tt}u^t, \;\mu = \sqrt{-p_\mu p^\mu}, \\
 L^2 &\equiv (g_{\theta\theta}u^\theta)^2 + \frac{L_z^2}{\sin^2\theta}=(u_\theta)^2 + \frac{(u_\phi)^2}{\sin^2\theta},\\
L_z &\equiv u_\phi=g_{\phi\phi}u^\phi,
\end{aligned}
\end{equation}
where $\varepsilon$ is the energy per unit mass, $L_z$ is the angular momentum per unit mass, 
and $L^2$ is the the square of the conserved total angular momentum per unit mass. Then the integration element
$d^4p$ can be changed to the volume element of four constants of the motion $d\varepsilon dL^2dL_zd\mu$. 

Assuming all DM particles in the system have the same rest mass $\mu_0$, then the mass distribution is a $\delta$ function.
Thus the distribution function can be simplified as $f^{(4)}(x,p) = \mu^{-3}f(\varepsilon,L^2,L_z)\delta(\mu-\mu_0)$. 
The element $d\mu$ can be integrated over. For later discussions, we just drop the subscript for simplicity. 
Considering the $\pm$ signs of $u^\mu$, we find that only the $t$ components of $J^\mu$ will not vanish in Schwarzschild case. 
We set $\theta=\pi/2$ to simplify further calculation thanks to the spherical symmetry of Schwarzschild metric. 
Finally the nonzero component of the mass current four-vector is
\begin{equation}\label{eqs1}
J_t(r)=-\frac{2}{r^2}\iiint d\varepsilon dL^2 dL_z \frac{\varepsilon f(\varepsilon,L^2,L_z)}{\sqrt{V(r)}\sqrt{L^2 - L_z^2}},
\end{equation}
where 
\begin{equation}
V(r) =  \varepsilon^2 - (1-\frac{2GM}{r})(1+\frac{L^2}{r^2}),
\end{equation}
and the density of DM spike is $\rho = \sqrt{-g^{tt}}\left|J_t\right|$.

We evaluate the phase space by the capture conditions of the black hole as in FRW~\cite{PhysRevD.96.083014}. It is assumed that only DM particles in bound orbits contribute to the density of spike, which means that we should take out the orbits which plunge in the black hole. 
We apply three constraints: $\varepsilon_{max} = 1$, $V(r)\ge 0$ and $U(\theta)\ge 0$. 
Here 
\begin{equation}
U(\theta) = L^2-\frac{L_z^2}{\sin^2\theta}.
\end{equation}
The potential $V(r)$ has a minimal value at unstable orbit $r_{unst}$, which hints that $dV(r)/dr = 0 $ and $d^2V(r)/dr^2 > 0$.
Imposing the positivity of $U(\theta)$, integration region $(x,y,z)$ can be restricted in the parameter space of $[0,1] \times [0,1] \times [0,1]$ as below,
\begin{equation}\label{boundary}
\begin{aligned}
\varepsilon(x) &=  x + (1-x)\varepsilon_{min},\\
L^2(y) &=  yL^2_{max} + (1-y)L^2_{crit},\\
L_z(y,z) &= (2z-1)\left|L(y)\right|.
\end{aligned}
\end{equation}
It allows the step size of the numerical integration to be fixed, thus facilitates our integration implementation.
By setting $L_z = \pm \sqrt{L^2}$ and $\varepsilon=1$ in $V(r)$, we obtain $L^2_{max}$. 
For critical orbits, it requires that $\varepsilon = \varepsilon_{min}$ and $L^2= L^2_{crit}$ at unstable orbit $r_{unst}$.
Meanwhile, $V(r)= 0$ has a double root.
These imply that we should solve a system of four polynomial equations of $(\varepsilon_{min},L^2_{crit},r_{unst},L_z^{\prime})$,
\begin{equation}
V(r)=0,U(\theta) = 0,\frac{dV(r)}{dr}(r=r_{unst}) = 0,V(r_{unst})=0.
\end{equation}
Now the integration boundaries are fully determined.
{Given the distance $r$, we obtain the approximate integral boundary by following the above steps. We sample in Eq.~\ref{boundary} to perform the integration of Eq.~\ref{eqs1} by Monte Carlo method. We accept those groups of parameter $(x,y,z)$ that satisfy $V(r)$ and $U(\theta) \ge 0$ to calculate the integrand in Eq.~\ref{eqs1}, for the rest we set the integrand to be zero. 
In this way, we can estimate the density $\rho$ at specific distance $r$.}

{Next, we use the conditions of adiabatic growth to solve the form of the distribution function $f$.}
We assume a spherically symmetric initial DM halo without a central black hole, 
which has a distribution function $f^{\prime}(E^{\prime},L^{\prime2},L^{\prime}_z)$ and gravitational potential $\Phi(r)$. 
Here $E^{\prime}$ is the classical energy per particle mass.
The adiabatic growth of the black hole in the halo keeps $f^{\prime}(E^{\prime},L^{\prime2},L^{\prime}_z) = f(\varepsilon,L^2,L_z)$, meaning the form of the distribution function is preserved. 

Through the adiabatic growth, the integrals of motion of the system are invariant. 
The adiabatic invariants consist of three components for $r, \theta$ and $\phi$.
Since both the initial dark halo and the Schwarzschild spacetime are spherically symmetric, matching $I^{\prime}_{\theta} = I_{\theta}$ and $I^{\prime}_{\phi} = I_{\phi}$, we obtain $L^{\prime}_z = L_z$ and $L^{\prime} = L$. The focus of the calculation is then about radial integral of motion.
The radial adiabatic invariant for a nonrelativistic DM particle is
\begin{equation}\label{eqs11}
I_r^{\prime}(E^{\prime},L^{\prime})=\oint dr \sqrt{2E^{\prime} - 2\Phi -\frac{{L^{\prime}}^2}{r^2}}.
\end{equation}
For the orbit of a DM particle in the Schwarzschild geometry, 
\begin{equation}\label{eqs12}
I_r(\varepsilon,L^2,L_z)=\oint u_r dr =\oint dr \frac{\sqrt{V(r)}}{1-2GM/r}.
\end{equation}
Later we shall firstly estimate $I^{\prime}_r = I_r$ for a Hernquist profile, which has the same $1/r$ behavior as NFW model in the center region.
\footnote{Although the distribution of DM in galactic halo might be not NFW like \cite{Salucci_2019}, we shall not consider such distributions in this paper.}
Hernquist profile has the density function of
\begin{equation}
\rho_H = \frac{\rho_0}{(r/r_s)(1+r/r_s)^3},
\end{equation}
with Newtonian gravitational potential
\begin{equation}
\Phi_H = -\frac{GM_{halo}}{r + r_s},
\end{equation}
where $r_s$ and $\rho_0$ are scale factors. $M_{halo} = 2\pi \rho_0 r_s^3$ is the total mass of DM halo.
For the Milky Way, we shall take $M_{halo}$ to be $10^{12}M_\odot$ and $r_s = 20~$kpc.
The distribution function of the profile was analytically derived by \cite{1990ApJ.356.359H}:
\begin{equation}
f_H(\tilde{\epsilon}) = \frac{M_{halo}}{\sqrt{2}(2\pi)^3(GM_{halo}r_s)^{3/2}}\tilde{f_H}(\tilde{\epsilon}),
\end{equation}
where  
\begin{equation}
\tilde{f_H}(\tilde{\epsilon}) = \frac{\sqrt{\tilde{\epsilon}}}{(1 - \tilde{\epsilon})^2} \left[ (1-2\tilde{\epsilon})(8{\tilde{\epsilon}}^2 - 2\tilde{\epsilon} - 3) + \frac{3\sin^{-1}{\sqrt{\tilde{\epsilon}}}}{\sqrt{\tilde{\epsilon}(1-\tilde{\epsilon})}}  \right]. 
\end{equation}
Here $\tilde{\epsilon}$ is a new dimensionless relative energy.
It is related to energy $E^{\prime}$ per unit particle mass or relativistic energy $\varepsilon$ per unit particle mass by 
$\tilde{\epsilon}^{\prime} = -r_s E^{\prime}/GM_{halo}$ in Newtonian case and
$\tilde{\epsilon} \equiv r_s(1-\varepsilon)/GM_{halo}$ in relativistic case, respectively.
Other dimensionless quantities are introduced below,
\begin{equation}\label{eqs17}
\tilde{M}  \equiv \frac{M}{M_{halo}},
x \equiv \frac{r}{r_s},
\tilde{L} \equiv \frac{L}{\sqrt{GM_{halo}r_s}},
\tilde{L}_z \equiv \frac{L_z}{GM_{halo}r_s},
\tilde{\psi} \equiv -\frac{r_s}{GM_{halo}}\Phi = \frac{1}{1+x}.
\end{equation}
Then we adopt the radial invariant for Hernquist profile and Schwarzschild geometry by substituting dimensionless relative energy $\tilde{\epsilon}$ and Eq.~\ref{eqs17} into Eq.~\ref{eqs11} and Eq.~\ref{eqs12}:
\begin{equation}\label{eqs18}
\begin{aligned}
I_r^H&= 2\sqrt{GM_{halo}r_s}\int_{x_{-}}^{x_{+}} dx \sqrt{-2\tilde{\epsilon}^{\prime} + \frac{2}{1+x} + \frac{{\tilde{L}}^2}{x^2}},\\
I_r^S&=2\sqrt{GM_{halo}r_s}\int_{x_{-}}^{x_{+}} dx \sqrt{-2\tilde{\epsilon} + 2\frac{\tilde{M}}{x} 
- \frac{\tilde{L}}{x^2} + \frac{{\tilde{\epsilon}}^2GM_{halo}}{r_s} + \frac{2GM_{halo}}{r_s} \frac{\tilde{M}{\tilde{L}}^2}{x^3}},
\end{aligned}
\end{equation}
where $x_{\pm}$ are the two positive roots of function in the square root which represent the turning points of the orbit.
For a set of $(x,y,z)$ in phase space, equating Eq.~\ref{eqs11} to Eq.~\ref{eqs12} gives $\tilde{\epsilon}$ and $\tilde{L}$. 
Because the radial invariant tends to be divergent as $\tilde{\epsilon} \rightarrow 0$, we adopt the operation similar to SFW, and rewrite the integrals of radial invariants into the domain $[0,1]$, to ensure the stability of numerical integration. Hence, using the relations $f_H(\tilde{\epsilon}) = f(\varepsilon,L^2,L_z) = f(x,y,z)$, the integration of $J_t(r)$ can be evaluated at any distance $r$.

Once the distribution function and adiabatic invariants are given, we can employ the above method to evaluate arbitrary DM spike densities near black holes. For instance, we consider a power-law initial dark halo $\rho_{p}(r) = \rho_0(r_s/r)^\gamma$, with $0 < \gamma < 2$. In terms of dark halo mass $M_{halo} = 4\pi \rho_0 r_s^3/(3-\alpha)$ and dimensionless quantities, we rewrite the phase space distribution and radial adiabatic invariant~\cite{PhysRevLett.83.1719},
\begin{equation}
\begin{aligned}
f(\tilde{\epsilon}) &= \frac{\rho_0 r_s^{3/2} (2-\gamma)^{3/2}}{(2\pi G M_{halo})^{3/2}}\frac{\Gamma(\beta)}{\Gamma(\beta-\frac{3}{2})}\left[-\frac{1}{(2-\gamma)\tilde{\epsilon}}\right]^{\beta},\\
I_r^p &= 2\sqrt{GM_{halo}r_s}\frac{B\left(\frac{1}{2-\gamma},\frac{3}{2}\right)}{2-\gamma}\left[ -\frac{\tilde{L}}{\lambda} + \sqrt{\frac{2}{2-\gamma}}
\left[-(2-\gamma)\tilde{\epsilon}\right]^\frac{4-\gamma}{2(2-\gamma)} \right], 
\end{aligned}
\end{equation}
where $\Gamma$ is Gamma function, $B$ is Beta function, $\beta = (6-\gamma)/[2(2-\gamma)]$ and $\lambda = [2/(4-\gamma)]^{1/(2-\gamma)}[(2-\gamma)/(4-\gamma)]^{1/2}$. 
Then we can use the above expressions to calculate the spike that grows from a power-law dark halo. 

\subsection{Velocity distribution of DM in Spike} \label{subsec:vd}
Above we have described how to calculate the total density of DM spike around a Schwarzschild black hole. It will be more useful if we can also get the velocity distribution of DM particles, which is needed in the estimation of dynamical friction. 

For given constants of motion $(\varepsilon,L^2,L_z)$, dividing the $r$, $\theta$ and $\phi$ components of the geodesics by the $t$ component, we obtain the three velocity components in spherical coordinates,
\begin{equation}
v^r = \frac{u^r}{u^t}, v^\theta = \frac{u^\theta}{u^t}, v^\phi = \frac{u^\phi}{u^t}.
\end{equation}
And the four velocity is written as $u^\mu = \gamma(1,v^j)$, where $v^j=u^j/u^t$ and $u^t = \gamma = (-g_{tt} - g_{ij}v^{i}v^{j})^{-1/2}$.
We obtain in Schwarzschild spacetime
\begin{equation}\label{eqs20}
\varepsilon^2 = -\frac{(-g_{tt})^2}{g_{tt}+v^2},
\end{equation}
where $v = \sqrt{g_{ij}v^{i}v^{j}}$. Then we can change the integration variable from $\epsilon$ to $v$ in Eq.~\ref{eqs1}, and obtain 
\begin{equation}\label{eqs2}
\frac{d\rho}{dv}=\frac{2v(-g_{tt})^{3/2}}{r^2(g_{tt}+v^2)^2}\iint dL^2 dL_z \frac{f(v,L^2,L_z)}{\sqrt{V(r)}\sqrt{L^2 - L_z^2}}.
\end{equation}
{Due to the spherical symmetry in Schwarzschild geometry, we expect Eq.~\ref{eqs2} gives an isotropic velocity distribution around the Schwarzschild black hole which depends on $r$ only. This is consistent with the analysis in Section.~\ref{subsec:growth} where we can calculate the density at specific distance $r$.} Eq.~\ref{eqs1} and Eq.~\ref{eqs2} give DM density and velocity distribution, respectively, 
and they are closely correlated. Later we shall calculate the density first and then take the derivative to get the distribution function.

{In the process of integrating Eq.~\ref{eqs1} with Monte Carlo method, at each sampling for Eq.~\ref{boundary}, the particle velocity can be calculated for a given set of sample $(x, y, z)$, 
which provides the access to construct the correlation between spike density and particle velocity.}
We limit the range of particle velocities with escape velocity
$v_{max} = \sqrt{2GM/r}$.
Apparently, $v_{max}$ is a function of $r$. 
So the upper bounds on velocity are different at various distances.
Therefore, we quantify the particle velocity by escape velocity $v_{max}(r)$ and define an unitized velocity $v/v_{max}$ to transform the region of value into $[0,1]$. 
Restriction can be imposed on the Monte Carlo integration process of mass current four-vector 
to count the DM particles at distance $r$ that contribute to the density of spike and have velocities below a certain value.
We call the result velocity distribution of density $\rho(r,<v/v_{max})$ in this case. 
Naturally, we obtain $\rho = 0$ as $v/v_{max}=0$ and $\rho_{max}(r) $ as $v/v_{max}=1$. 
We parameterize DM's isotropic velocity distribution in three-dimensions velocity phase space as following
\begin{equation}\label{density-v}
\rho(r, < \alpha = v/v_{max})=\int_{0}^{\alpha} 4\pi a^2 f'(r,a)da,
\end{equation}
By differentiating the density in Eq.~\ref{density-v} by quantified velocity $\alpha=v/v_{max}$, we redefine the velocity distribution function,
\begin{equation}
f(r,\alpha = v/v_{max})=\frac{1}{\rho(r,1)}\frac{d\rho(r,\alpha)}{d\alpha}  
=\frac{4\pi \alpha^2 f'(r,\alpha)}{\rho(r,1)}.
\end{equation}

Given the distance $r$ and velocity $v$ to infer the value of $\alpha$, $\rho(r,\alpha)$ represents the density constructed by the DM particles whose velocities are below $v$. We will present the results under various parameter conditions to intuitively reflect the accumulation of DM particles around black hole in Section \ref{sec:RE}.
In the following Section \ref{sec:GW}, we elucidate the DF effects of DM density with velocity distribution on the orbiting objects in EMRIs and investigate the impact on the GW from EMRIs.

\section{Effects on Gravitational Waves from EMRIS} \label{sec:GW}
Now we are in a position to consider the effects on EMRI from a DM spike. We shall include the DF of DM as well as 3PN gravitational radiation 
reaction~\cite{10.1093/ptep/ptv092} to obtain the orbital energy loss and the evolution of the constants of motion. 
Introducing the evolving constants of motion, we construct numerical kludge waveform~\cite{PhysRevD.75.024005} to compute the orbital trajectory of orbiting body. 
Then we calculate the GW waveform with or without DF.

\subsection{Gravitational radiation and Dynamical Friction} \label{subsec:DF}
In the Schwarzschild spacetime, the orbit stays in a plane and consequently we can obtain the orbital trajectory $(r(t),\phi(t))$ of test body by integrating the geodesic equations,
\begin{equation}\label{eqs21}
\begin{aligned}
\frac{dt}{d\tau} &= \frac{E}{1-2GM/r},\\
\frac{dr}{d\tau} &= \pm \sqrt{E^2 - \left(1-\frac{2GM}{r}\right)\left(1+\frac{L^2}{r^2}\right)},\\
\frac{d\phi}{d\tau} &= \frac{L_z}{r^2}.
\end{aligned}
\end{equation}
Numerical integration can be optimized by defining
\begin{equation}\label{eqs22}
r = \frac{p}{1+e\cos\psi}.
\end{equation}
Here $p$ is semi-latus rectum and $e$ is orbital eccentricity. Then we work with $\psi$ and $\phi$, instead of $r$ and $\phi$. 
We combine Eq.~\ref{eqs21} and Eq.~\ref{eqs22}, then obtain the equation for $\psi$,
\begin{equation}\label{eqs111}
\frac{d\psi}{dt} = \frac{\sqrt{(1-E^2)p(p-er_{in}(1+e\cos{\psi})}}{E\sqrt{1-e^2}r^3/(r-2GM)},
\end{equation}
where $r_{in}$ is one of the zero points of the equation $dr/d\tau$, which is located inside the horizon.
The other two zero points are the turning points of the orbit, expressed by $r_a=p/(1-e)$ and $r_p=p/(1+e)$.
The evolution equation Eq.~\ref{eqs111} corresponds to the Schwarzschild limit of the result of \cite{PhysRevD.75.024005}.

Taking GW radiation reaction into account, the orbital constants dissipate over time,
\begin{equation}\label{eqs24}
\left(\frac{dE}{dt}\right)_{\textrm{GW}} = g_E(m,M,p,e),
\left(\frac{dL_{z}}{dt}\right)_{\textrm{GW}} = g_{L_z}(m,M,p,e).
\end{equation}
We consider the 3PN back-reaction of the orbital constants \cite{10.1093/ptep/ptv092}. In this way, gravitational radiation is introduced into geodesic motion. Moreover, orbiting objects are dragged by the gravity of the surrounding matter, causing energy loss. Since the velocities of DM particles near black holes are close to the speed of light $c$, we consider relativistic DF force \cite{10.1111/j.1365-2966.2007.12408.x,PhysRevD.104.103014} described by
\begin{equation}\label{eqs25}
\left(\frac{dE}{dt}\right)_{\textrm{DF}} = F_{\textrm{DF}}\cdot v = -\frac{4\pi \rho(r,<v) G^2 m^2 \gamma^2 [1+(v/c)^2]^2}{v}\ln\Lambda,
\end{equation}
where $\gamma = 1/\sqrt{1-(v/c)^2}$ is the Lorentz factor, and $\ln\Lambda$ is the Coulomb logarithm adopted as 10. In this framework, the energy loss from gravitational drag is contributed by the DM particles whose velocities are below the orbiting relativistic compact object. Similarly, the angular momentum loss rate is
\begin{equation}\label{eqs26}
\left(\frac{dL_z}{dt}\right)_{\textrm{DF}} = r\cdot F_{\textrm{DF}}\frac{r\dot{\psi}}{v}= -\frac{4\pi \rho(r,<v) G^2 m^2 r^2 \dot{\psi} \gamma^2 [1+(v/c)^2]^2}{v^2}\ln\Lambda.
\end{equation}
Put the above together, we have the evolution equations for energy and angular momentum,
\begin{equation}\label{eqs27} 
\frac{dE}{dt}=\left(\frac{dE}{dt}\right)_{\textrm{GW}} + \left(\frac{dE}{dt}\right)_{\textrm{DF}},\;
\frac{dL_z}{dt}= \left(\frac{dL_z}{dt}\right)_{\textrm{GW}} + \left(\frac{dL_z}{dt}\right)_{\textrm{DF}}.
\end{equation}
Combining with the geodesic equation and solving numerically, we obtain the trajectory of the orbiting object, which is used to calculate the corresponding gravitational waveform. 

\subsection{Gravitational Wave} \label{subsec:NK}
Once we have solved the equations of motion for the orbit, we proceed to calculate the corresponding gravitational waves. Using quadrupole-octupole formula \cite{PhysRevD.15.965}, we have the wave field
\begin{equation}\label{eqs28}
h_{jk}(t,\Vec{x}) =\frac{2G}{c^4}\frac{1}{d}[\ddot{I}_{jk}-2n_i\ddot{S}_{ijk}+n_i\dddot{M}_{ijk}]\big{|}_{t'=t-\frac{d}{c}},
\end{equation}
where ${I}_{jk}$ is the mass quadrupole moment, ${S}_{ijk}$ is the current quadrupole moment, ${M}_{ijk}$ is the mass octupole moment, 
an overdot denotes a time-derivative, $d$ is the distance from the source to detector, and $n_i$ are the components of direction vector $\hat{\mathbf{n}}=\mathbf{d}/d=(\sin\theta_D\cos\phi_D,\sin\theta_D\sin\phi_D,\cos\phi_D)$, specified by the azimuth $\phi_D$ and latitude $\theta_D$ of the source in the sky. 
In the extreme-mass-ratio limit for point mass $m$ (the mass ratio of binary 
is $m/M \ll 1$), the multipole moments are 
\begin{equation}\label{eqs29}
\begin{aligned}
{I}_{jk} &= mx_{j}x_{k},\\
{S}_{ijk}&= mv_{i}x_{j}x_{k},\\
{M}_{ijk}&= mx_{i}x_{j}x_{k}.
\end{aligned}
\end{equation}
Here $x_{i,j,k}$ correspond to the components of the position in Cartesian coordinates $(x,y,z)$ and $v_{i} = dx_{i}/dt$. 
Here and hereafter, $(m,M,d)$ of the source are referred as the redshifted masses and luminosity distance. 

Projecting the waveform in the transverse-traceless (TT) gauge, we have the plus $(+)$ and the cross $(\times)$ polarizations of waveform, $h_+$ and $h_\times$, 
\begin{eqnarray}\label{eqs30}
h_{jk}^{TT} & = &
 \left(
\begin{array}{ccc}
0   &   0 &   0\\
0   &   h_{+} &   h_{\times}\\
0   &   h_{\times} &   -h_{+}
\end{array}     \right) .
\end{eqnarray}
When taking into account the responses of space-based GW detectors like LISA and Taiji, it is convenient to consider the strain $h_{I,II}$, which are linear combinations of $h_+$ and $h_{\times}$,
\begin{equation}\label{eqs32}
h_{I} = \frac{\sqrt{3}}{2}(F_{I}^+ h_{+} + F_{I}^{\times}h_{\times}),
h_{II} = \frac{\sqrt{3}}{2}(F_{II}^+h_{+} + F_{II}^{\times}h_{\times} ),
\end{equation}
where $F^{+,\times}$ are antenna pattern functions \cite{PhysRevD.49.6274}. The polarization angle of the source $\psi_D$ in $F^{+,\times}$ is given by
\begin{equation}\label{eqs34}
\psi_D = \arctan\left(\frac{\hat{\mathbf{L}}\cdot\hat{\mathbf{z}}-(\hat{\mathbf{L}}\cdot\hat{\mathbf{n}})(\hat{\mathbf{z}}\cdot\hat{\mathbf{n}})}{\hat{\mathbf{n}}\cdot(\hat{\mathbf{L}}\times\hat{\mathbf{z}})}\right).
\end{equation}
$\psi_D$ depends on the unit vector $\hat{\mathbf{L}}$ of orbital angular momentum $\mathbf{L}$ and the basis vector $\hat{\mathbf{z}}$ of the z-axis in detector-based coordinate system. 
For a orbit in equatorial plane, angular momentum $\mathbf{L}$ is conserved. Thus the unit vector $\hat{\mathbf{L}}$ is a constant vector and $\psi$ is constant. For illustration, we set $\psi_D = \pi/2$ ($\hat{\mathbf{L}} = \hat{\mathbf{z}}$) and $(\theta_D,\phi_D)=(\pi/2,0)$ in later calculations. For the strain $h(t) = h_{I} +ih_{II}$ in time domain, we only show $h_{I}$ because $h_{I}$ and $h_{II}$ have similar features.

\begin{table*}[ht!]
\caption{\label{tab:parameter}%
Physical parameters of three cases. The initial parameters of the three cases we considered. We fit the profiles with Eq.~\ref{eqs35}, and present the best fitted parameters in the last column.}
\begin{ruledtabular}
\begin{tabular}{cccccc}
 Case&Black hole mass&Initial halo&Halo parameters
&Index&Fitting parameters\\
 & ($M$) & (Type) & $(M_{halo},r_s)$ &($\gamma$) & ($\kappa$, $\eta$ , $\omega$)\\ \hline
 C1 & $4.6\times10^6M_\odot$ & Hernquist & $(10^{12}M_\odot,20~\mathrm{kpc})$ & - & ($5.33\times10^{20}~\mathrm{GeV/cm^3}$, $1.99$, $2.07$) \\
 C2 & $10^4M_\odot$ & Hernquist & $(4.5\times10^{8}M_\odot,1.85~\mathrm{kpc})$ & - & ($6.15\times10^{24}~\mathrm{GeV/cm^3}$, $2.03$, $2.11$) \\
 C3 & $10^4M_\odot$ & Power law & $(7.3\times10^{8}M_\odot,1.85~\mathrm{kpc})$ & 7/4 & ($5.83\times10^{26}~\mathrm{GeV/cm^3}$, $2.04$, $2.16$) \\
\end{tabular}
\end{ruledtabular}
\end{table*}

\section{Results and Discussions} \label{sec:RE}

Now we demonstrate the density profiles of DM spikes growing from different initial halos around black holes of several sizes. We consider three cases for comparison. The physical parameters for each case are listed in Table.~\ref{tab:parameter}. {Specifically, we consider a Hernquist dark halo around the black hole at the center of the Milky Way galaxy (C1), a Hernquist dark halo around a black hole with the mass $M=10^4M_\odot$ (C2) and a power-law dark halo around a black hole with the mass $M=10^4M_\odot$(C3) in a dwarf galaxy.}

Note that the DM halo mass $M_{halo}$ and the scale $r_s$ are correlated. For the case with black hole mass $M=10^4M_\odot$, we follow the approach of \cite{PhysRevD.99.043533} and \cite{PhysRevD.102.103022} to relate the halo parameters $\rho_0$ and $r_s$ to central black hole mass $M$ by mass-velocity-dispersion relation. Thus we can estimate the density of the DM spike around a black hole. For example, for $M = 10^4M_\odot$, we obtain the halo scale factor $r_s=1.85~$kpc and halo mass $M_{halo} \approx 4.5\times10^8M_\odot$. 
The halo mass for Hernquist profile is determined by $M_{halo} = 2\pi\rho_0 r_s^3$, while halo with power-law profile $M_{halo} \approx 7.3\times10^8M_\odot$ with scale $r_s = 1.85 \mathrm{kpc}$. Here the power-law index is $\gamma = 7/4$.
\begin{figure*}[ht!]
\includegraphics[width=3in]{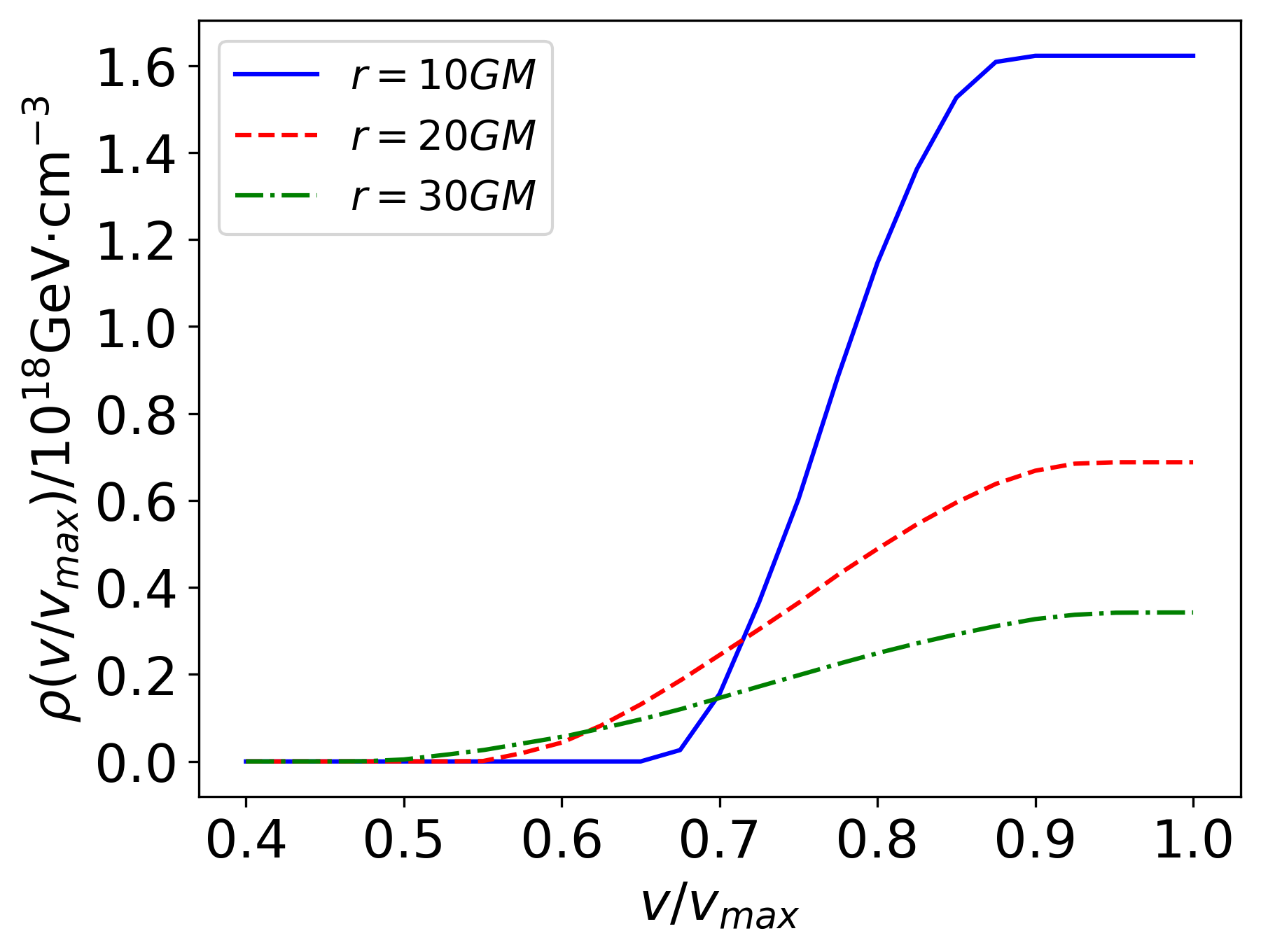}{(a)}
\includegraphics[width=3in]{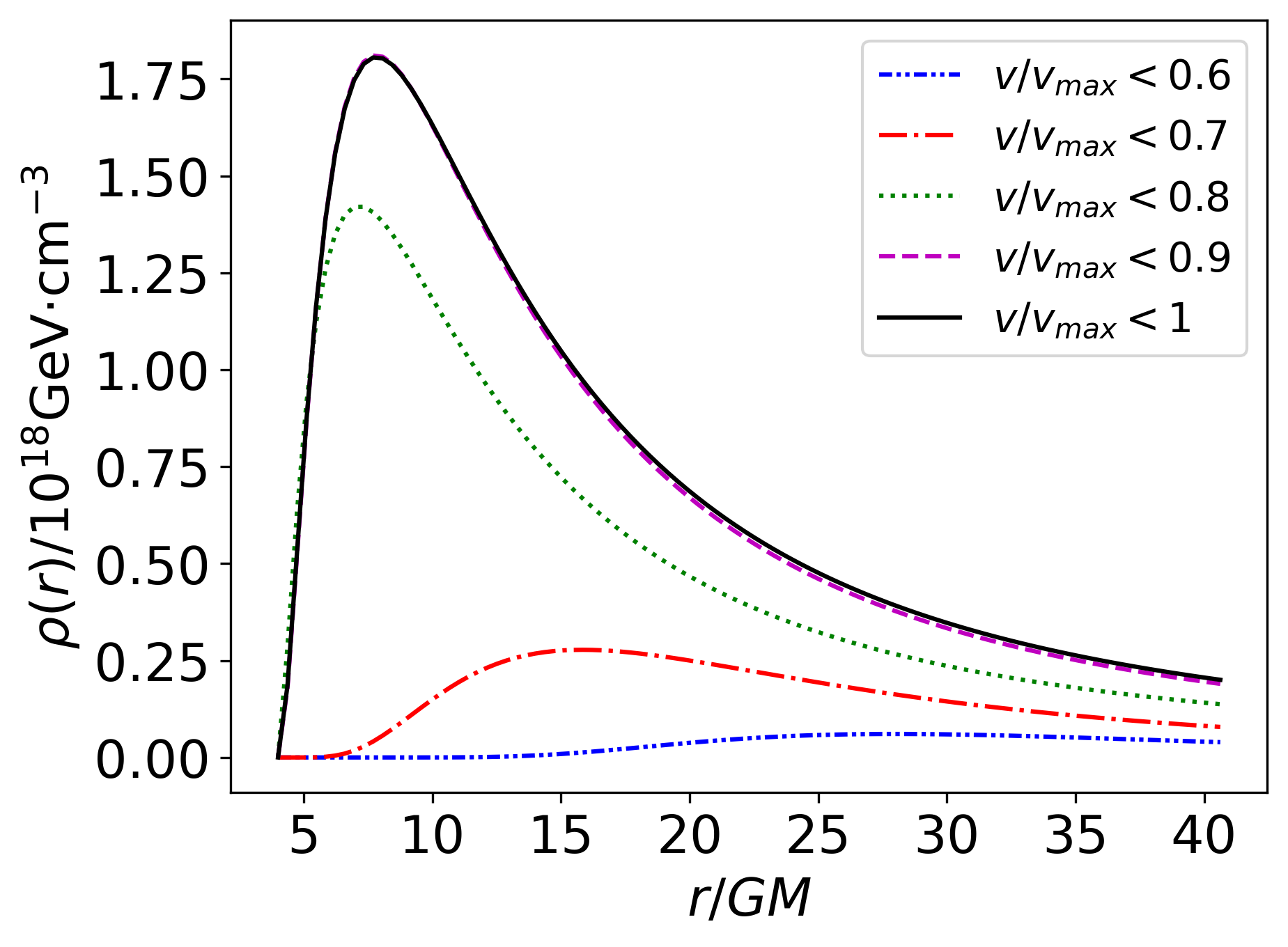}{(b)}
\includegraphics[width=3in]{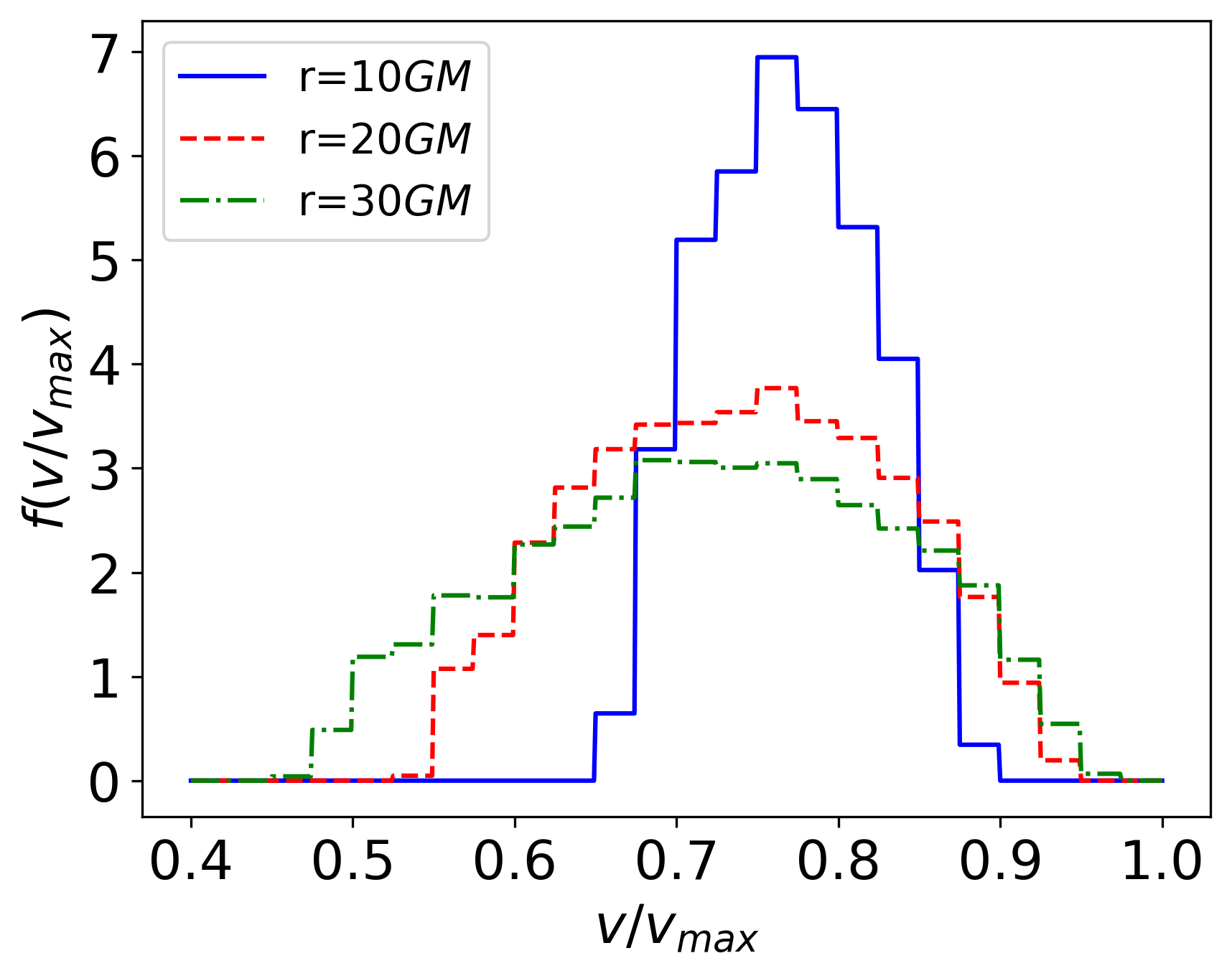}{(c)}
\includegraphics[width=3in]{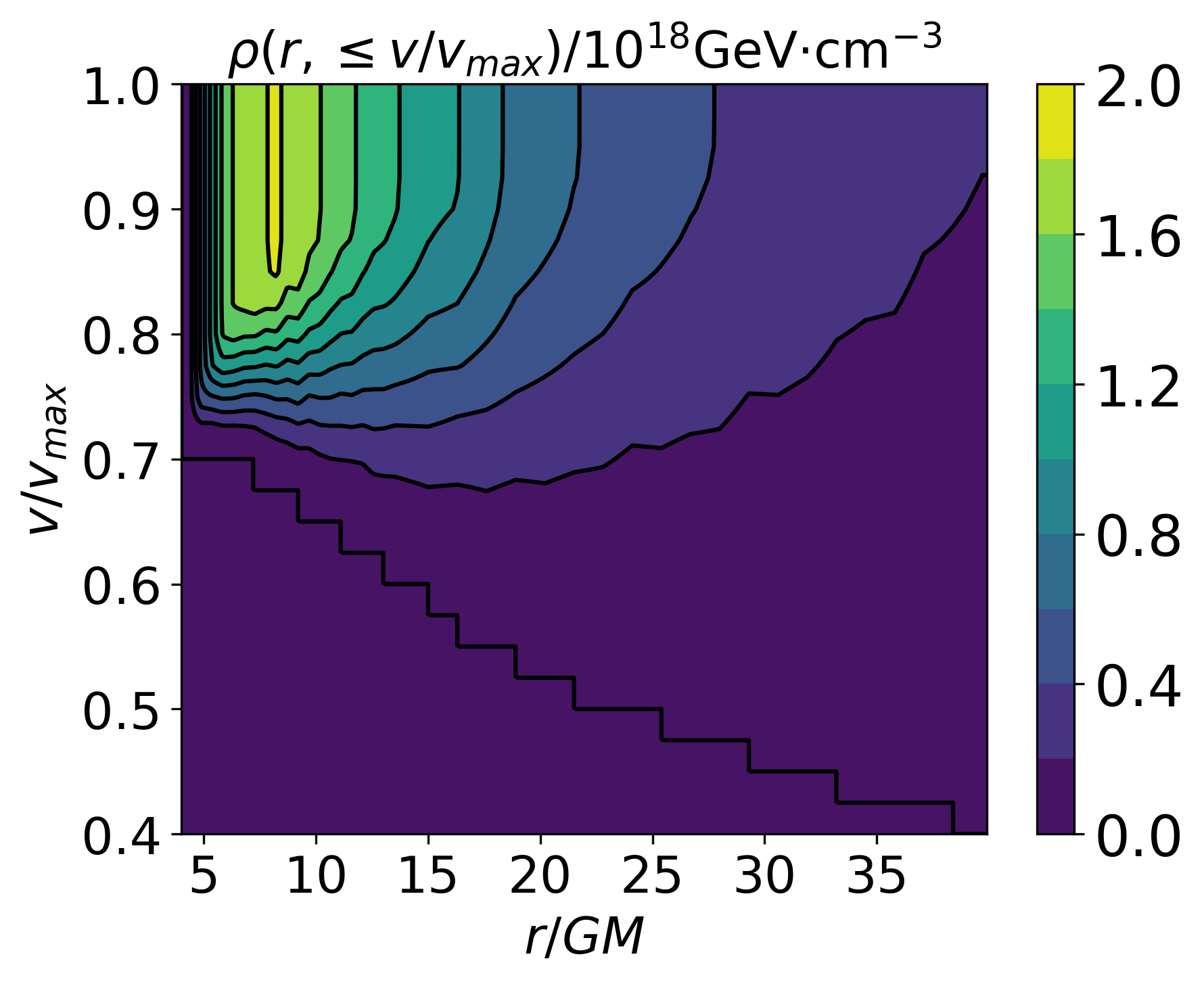}{(d)}
\caption{\label{fig:1} 
DM distribution around the black hole in Milky Way. Dark halo parameters $(M_{halo}, r_s)$ are $(10^{12}M_\odot, 20\mathrm{kpc})$. (a) The density-velocity relation $\rho(v/v_{max})$ of DM at specific distances. (b) The density-distance relation $\rho(r)$ of DM below various velocities. (c) The DM distribution $f(v/v_{max})$ of velocity at specific distances. (d) DM density $\rho(r,\le v/v_{max})$ in the $r-v/v_{max}$ plane.}
\end{figure*}

\begin{figure*}[ht!]
\includegraphics[width=3in]{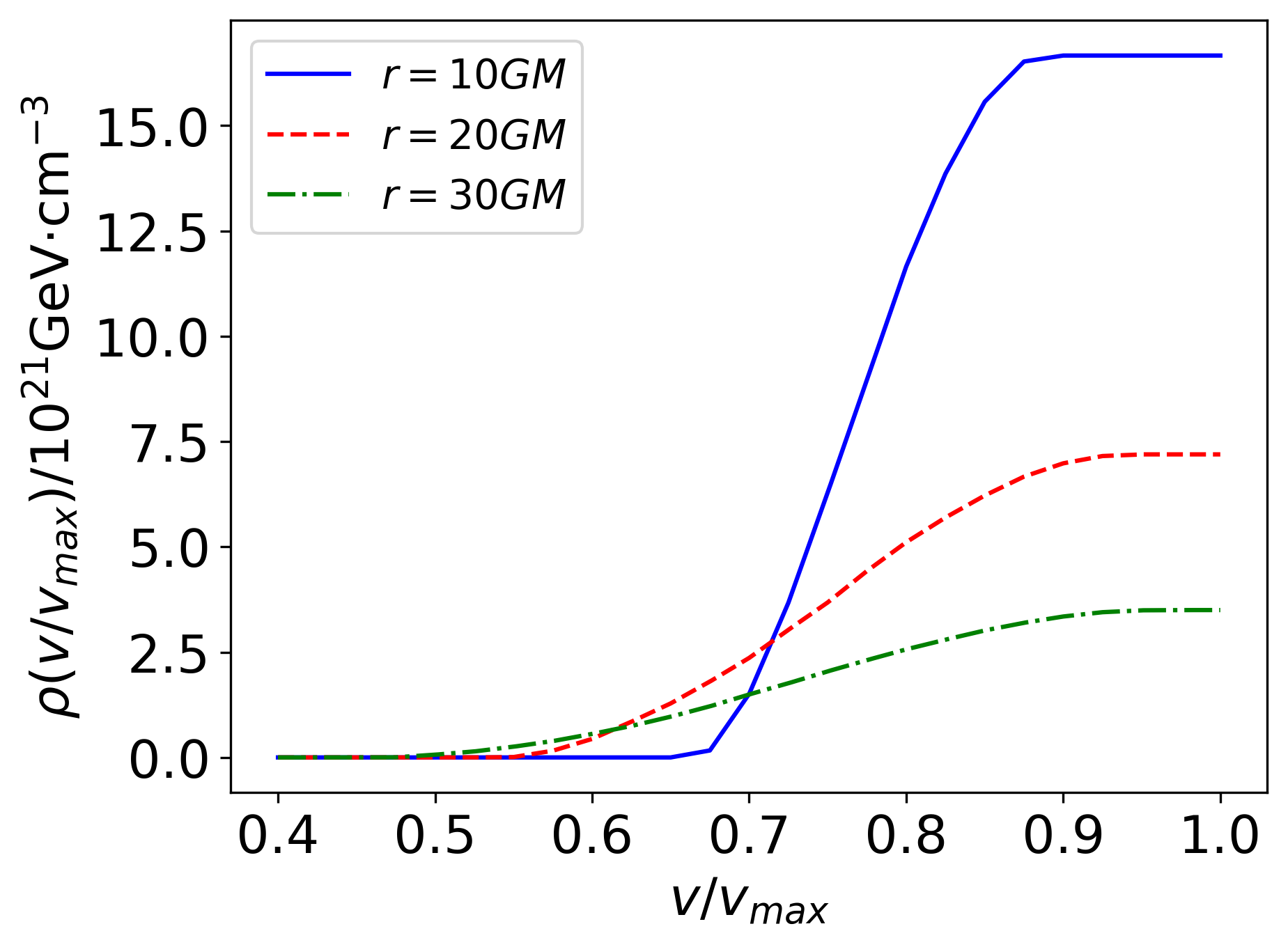}{(a)}
\includegraphics[width=3in]{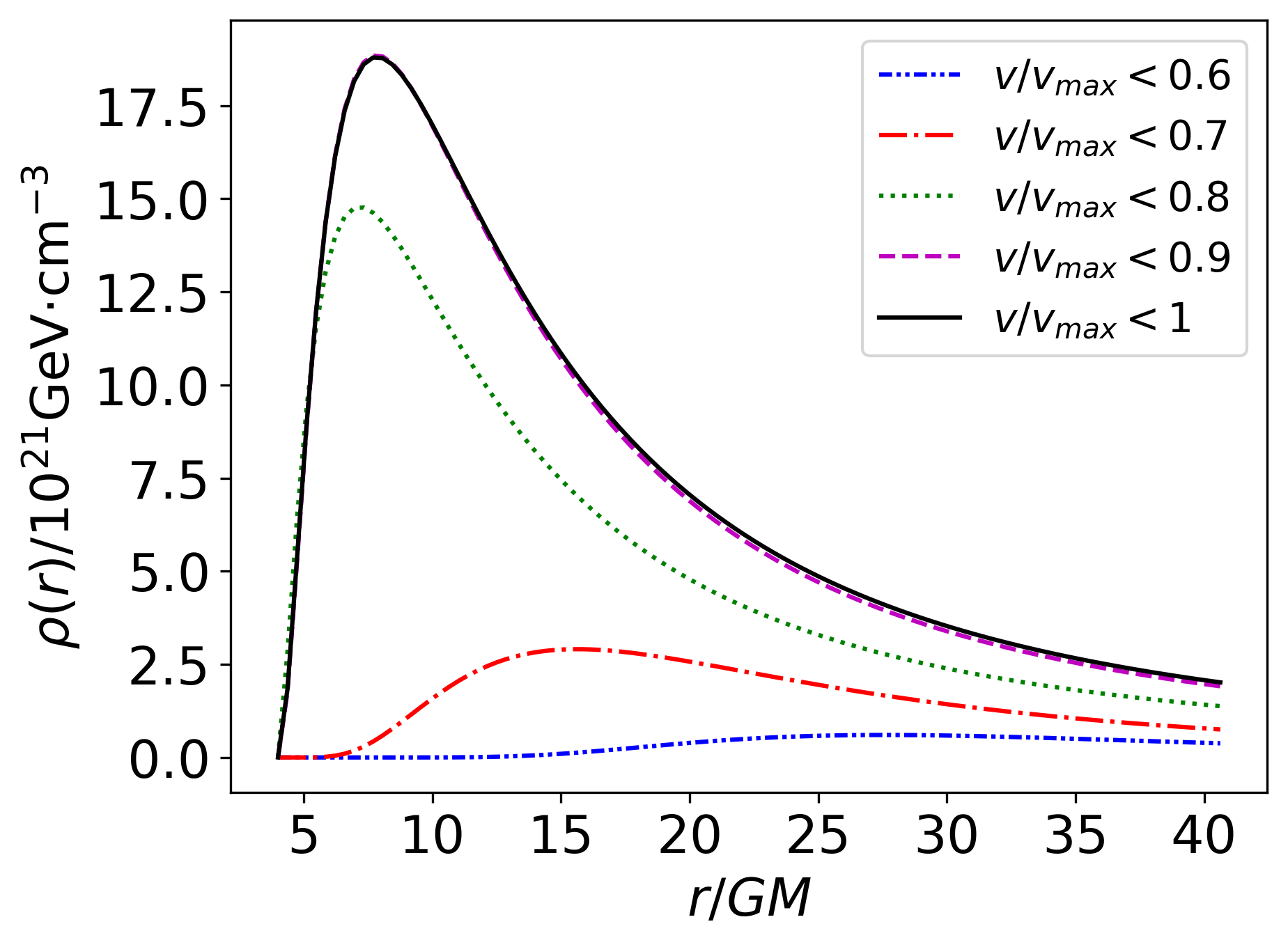}{(b)}
\includegraphics[width=3in]{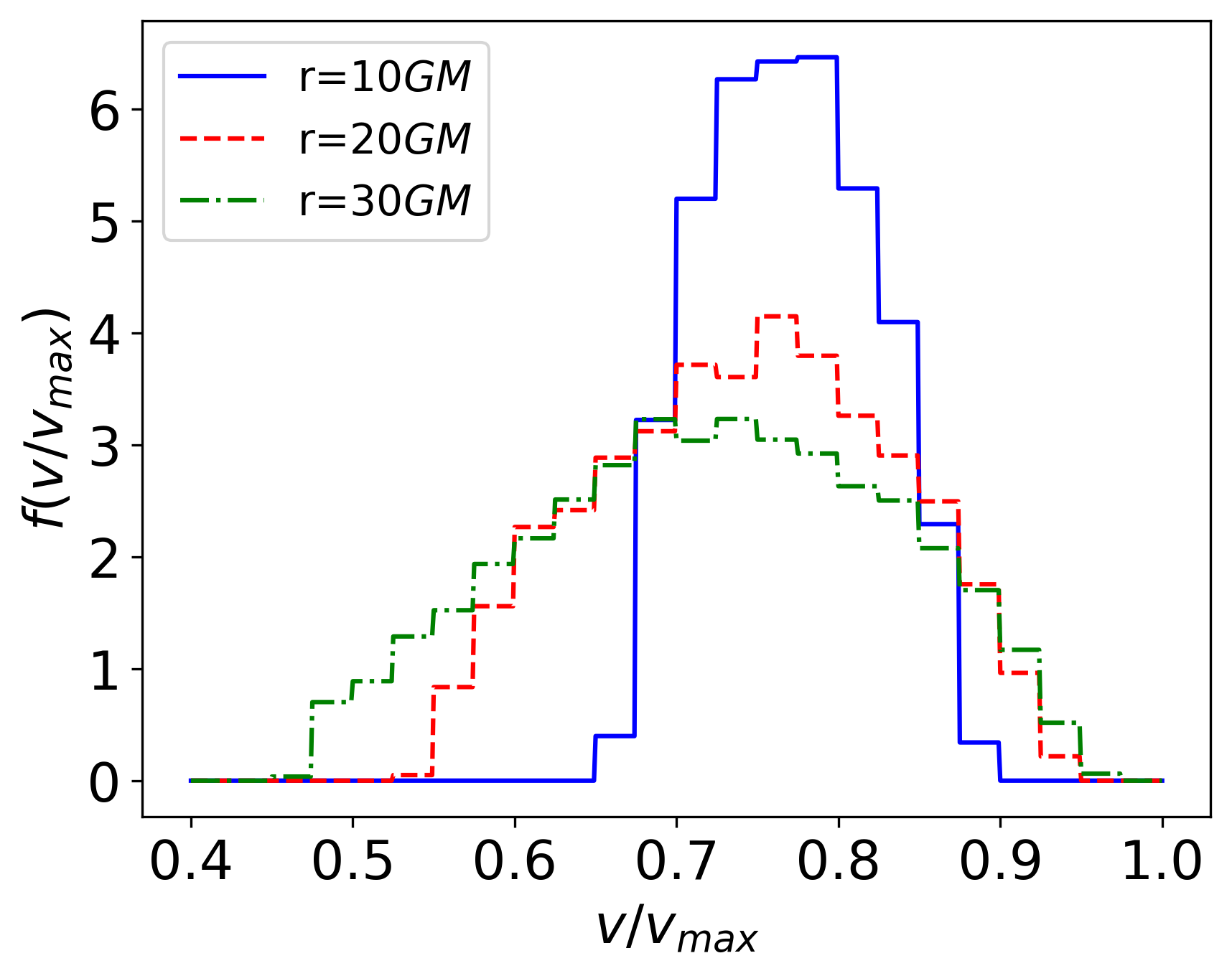}{(c)}
\includegraphics[width=3in]{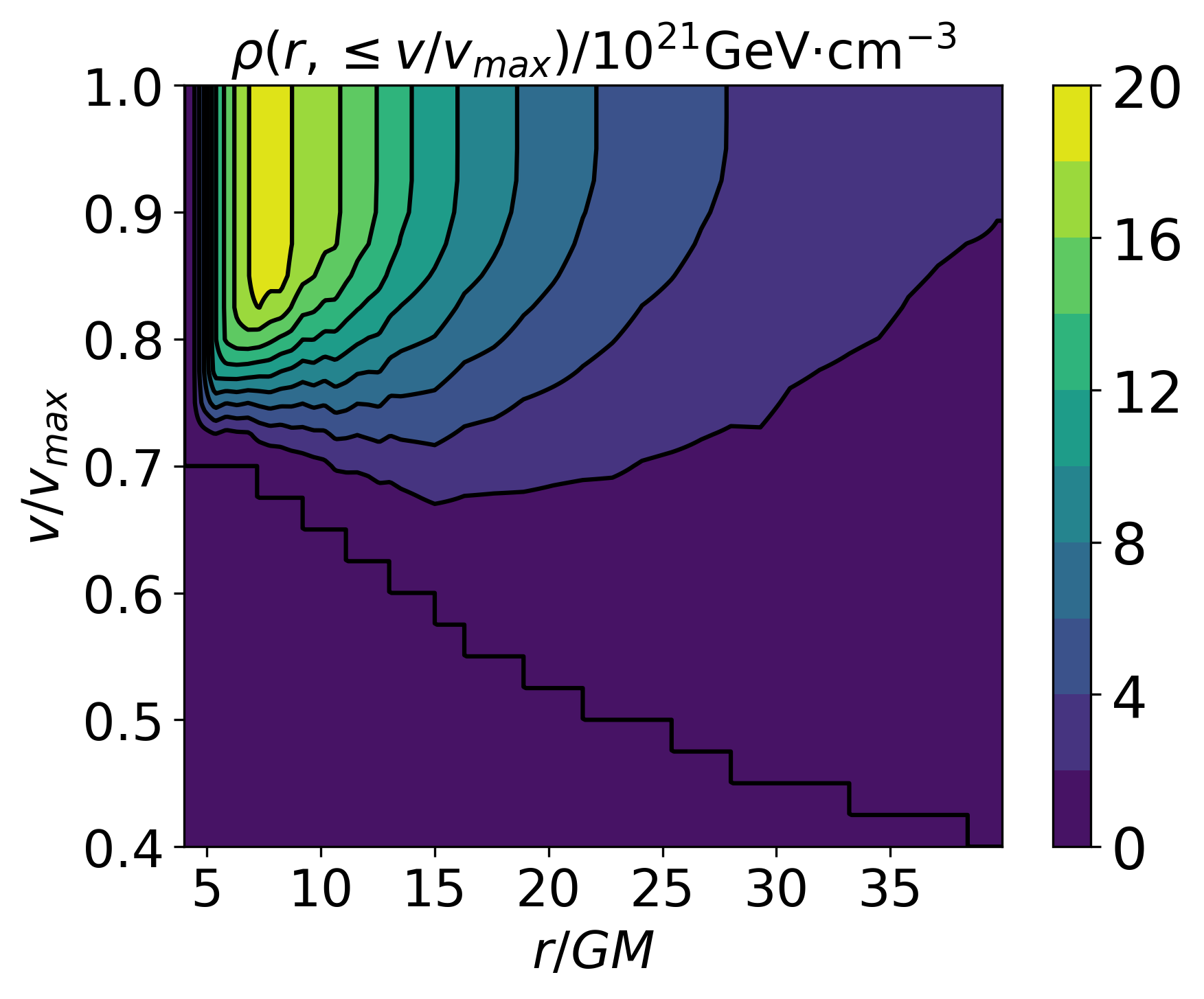}{(d)}
\caption{\label{fig:2} 
DM distribution around a black hole with mass $M = 10^4M_\odot$ with an initial Hernquist profile. Dark halo parameters $(M_{halo}, r_s)$ are $(4.5\times10^{8}M_\odot, 1.85\mathrm{kpc})$. (a) The density-velocity relation $\rho(v/v_{max})$ of DM at specific distances. (b) The density-distance relation $\rho(r)$ of DM below various velocities. (c) The DM distribution $f(v/v_{max})$  of velocity at specific distances. (d) DM density $\rho(r,\le v/v_{max})$ in the $r-v/v_{max}$ plane.}
\end{figure*}

\begin{figure*}[ht!]
\includegraphics[width=3in]{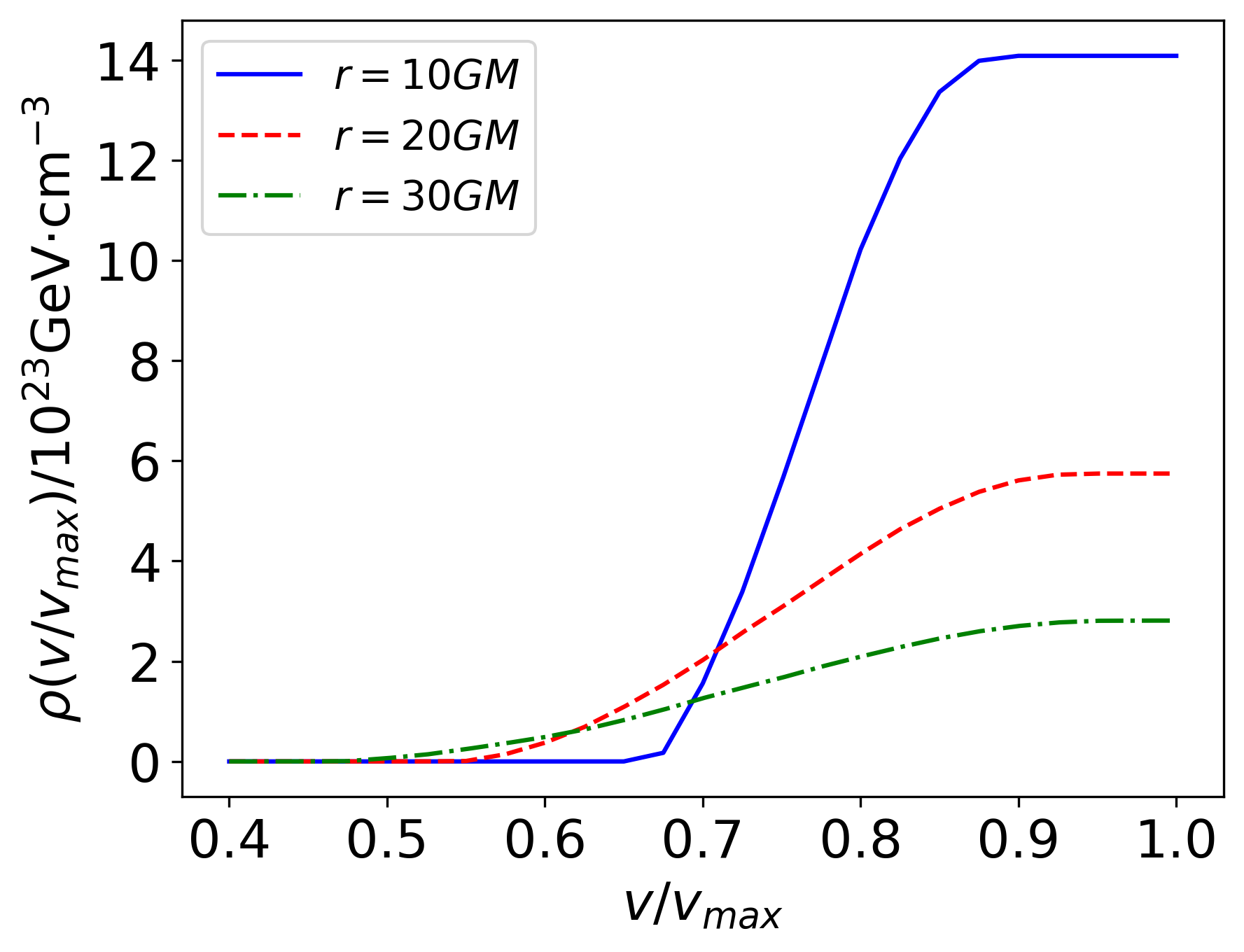}{(a)}
\includegraphics[width=3in]{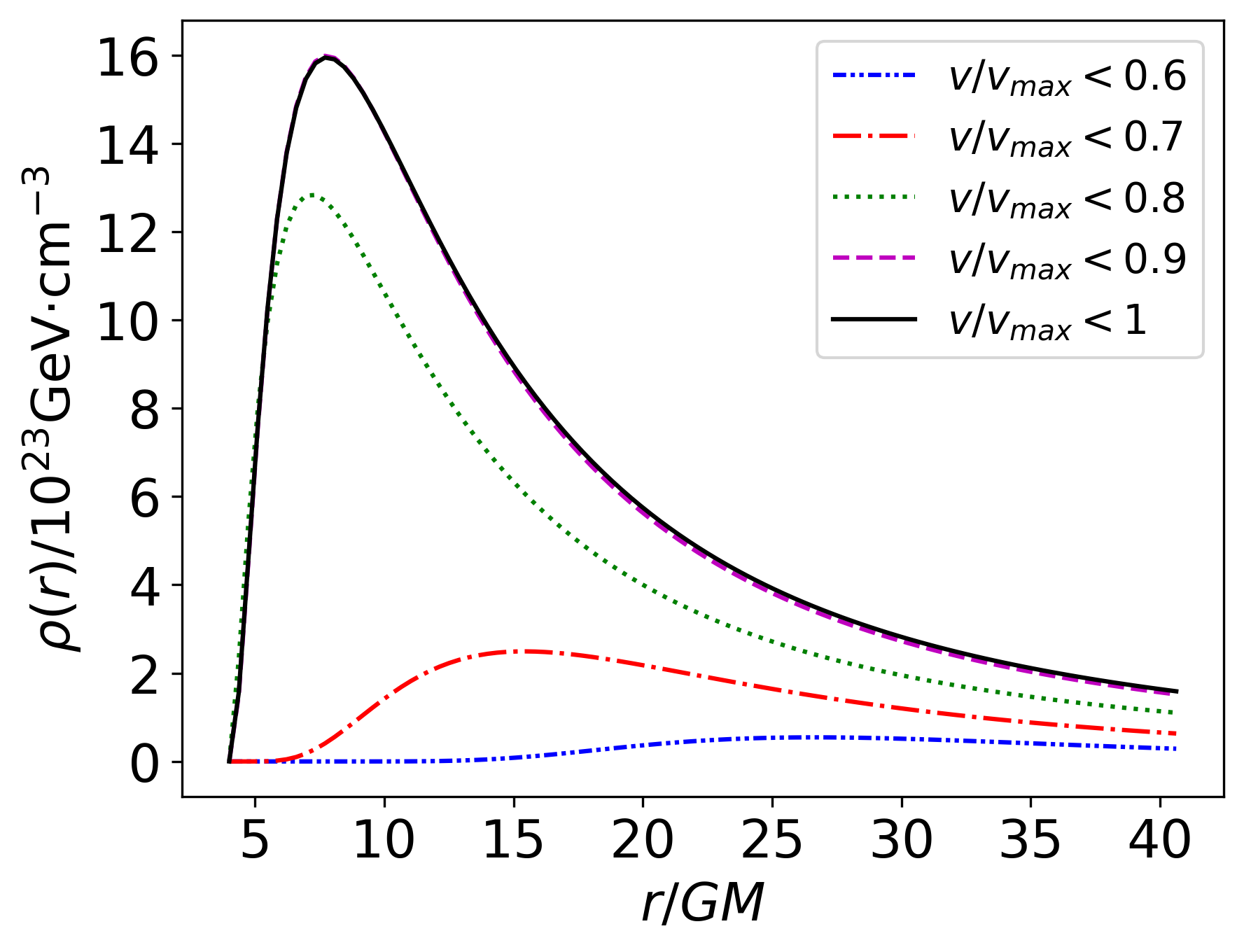}{(b)}
\includegraphics[width=3in]{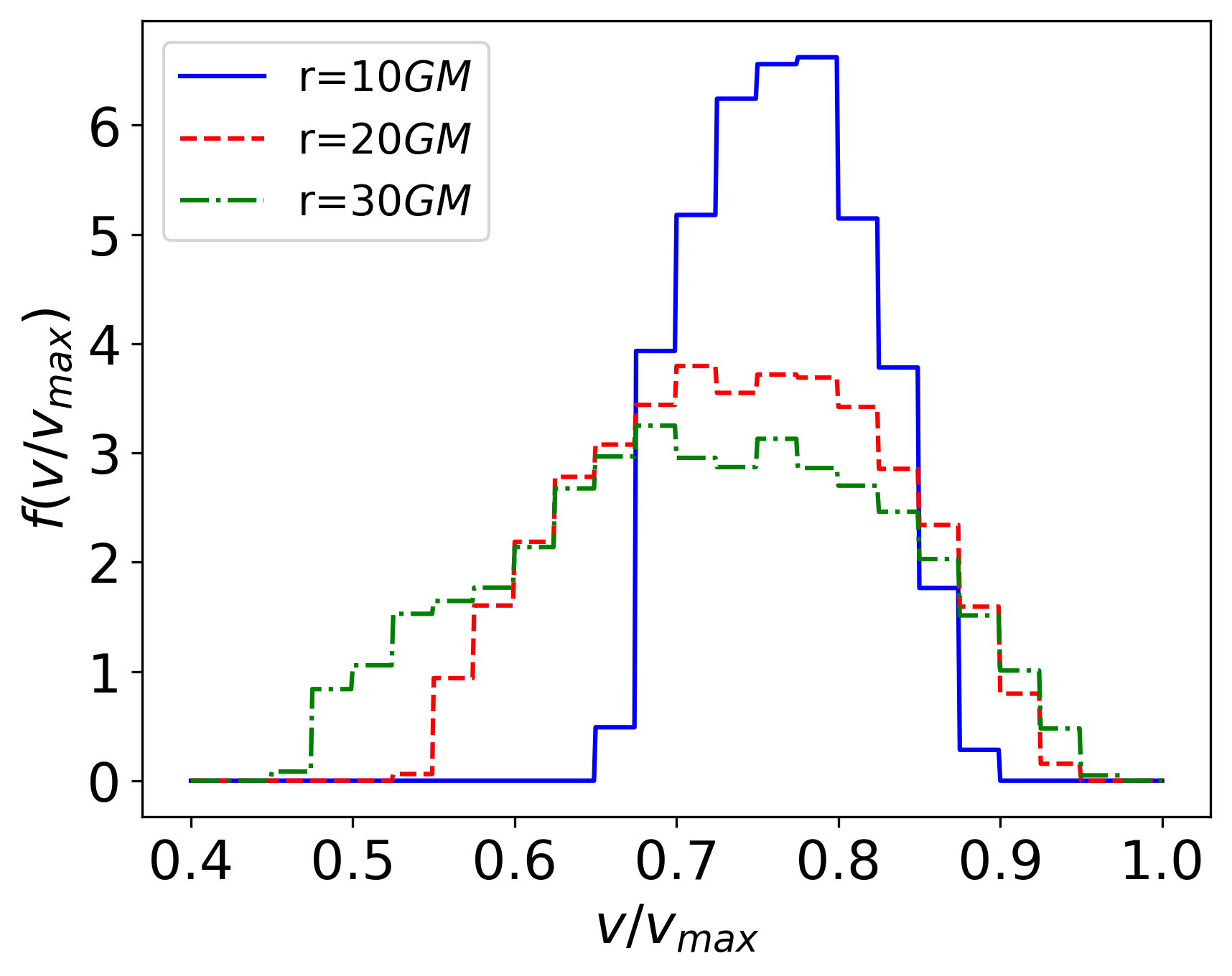}{(c)}
\includegraphics[width=3in]{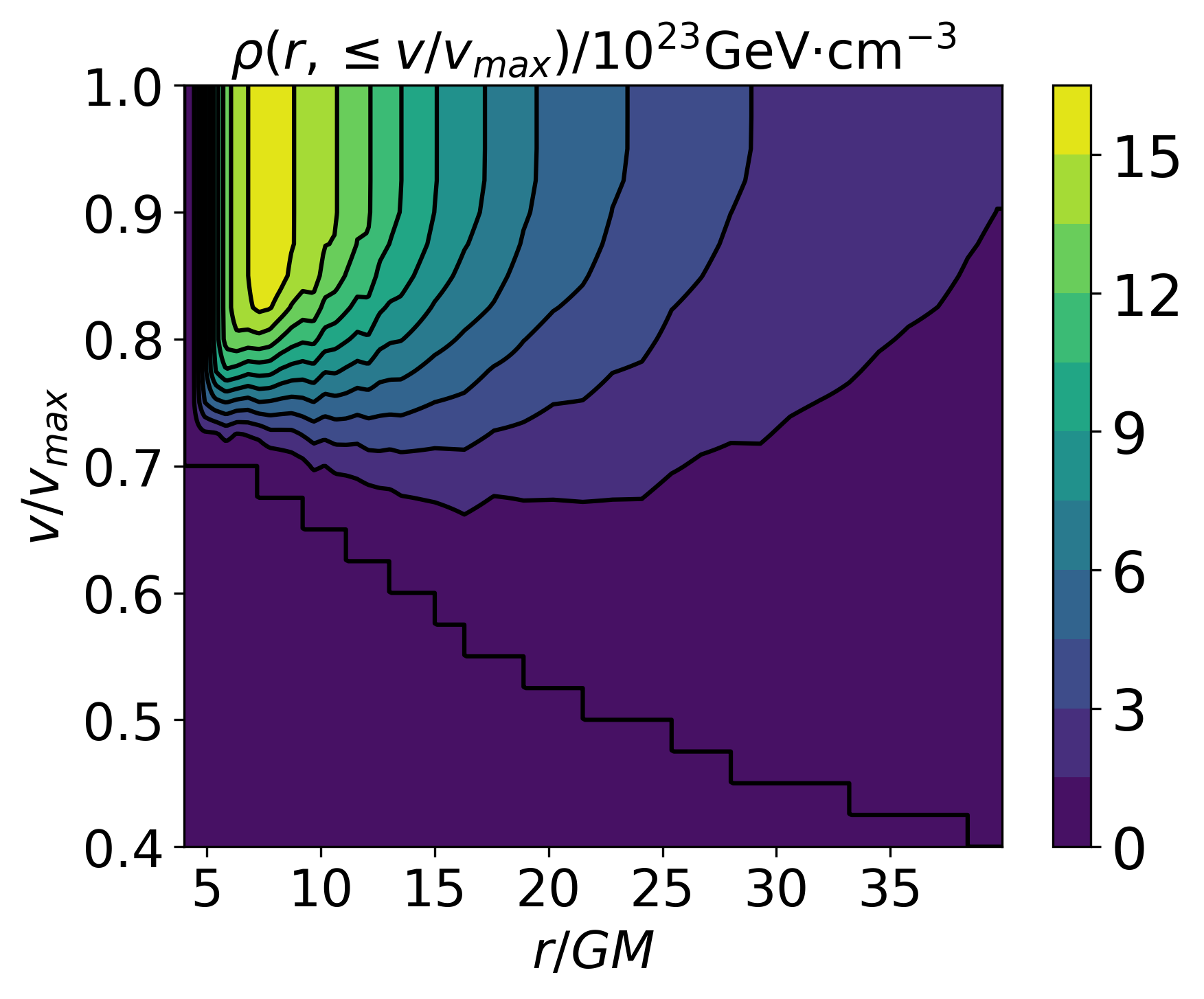}{(d)}
\caption{\label{fig:3} 
DM distribution around a center black hole with mass $M = 10^4M_\odot$ with an initial power-law profile. Dark halo parameters $(M_{halo}, r_s, \gamma)$ are $(7.3\times10^{8}M_\odot, 1.85\mathrm{kpc}, 7/4)$. (a) The density-velocity relation $\rho(v/v_{max})$ of DM at specific distances. (b) The density-distance relation $\rho(r)$ of DM below various velocities. (c) The DM distribution $f(v/v_{max})$ of velocity at specific distances. (d) DM density $\rho(r,\le v/v_{max})$ in the $r-v/v_{max}$ plane.}
\end{figure*}

\begin{figure*}[ht!]
\includegraphics[width=6in]{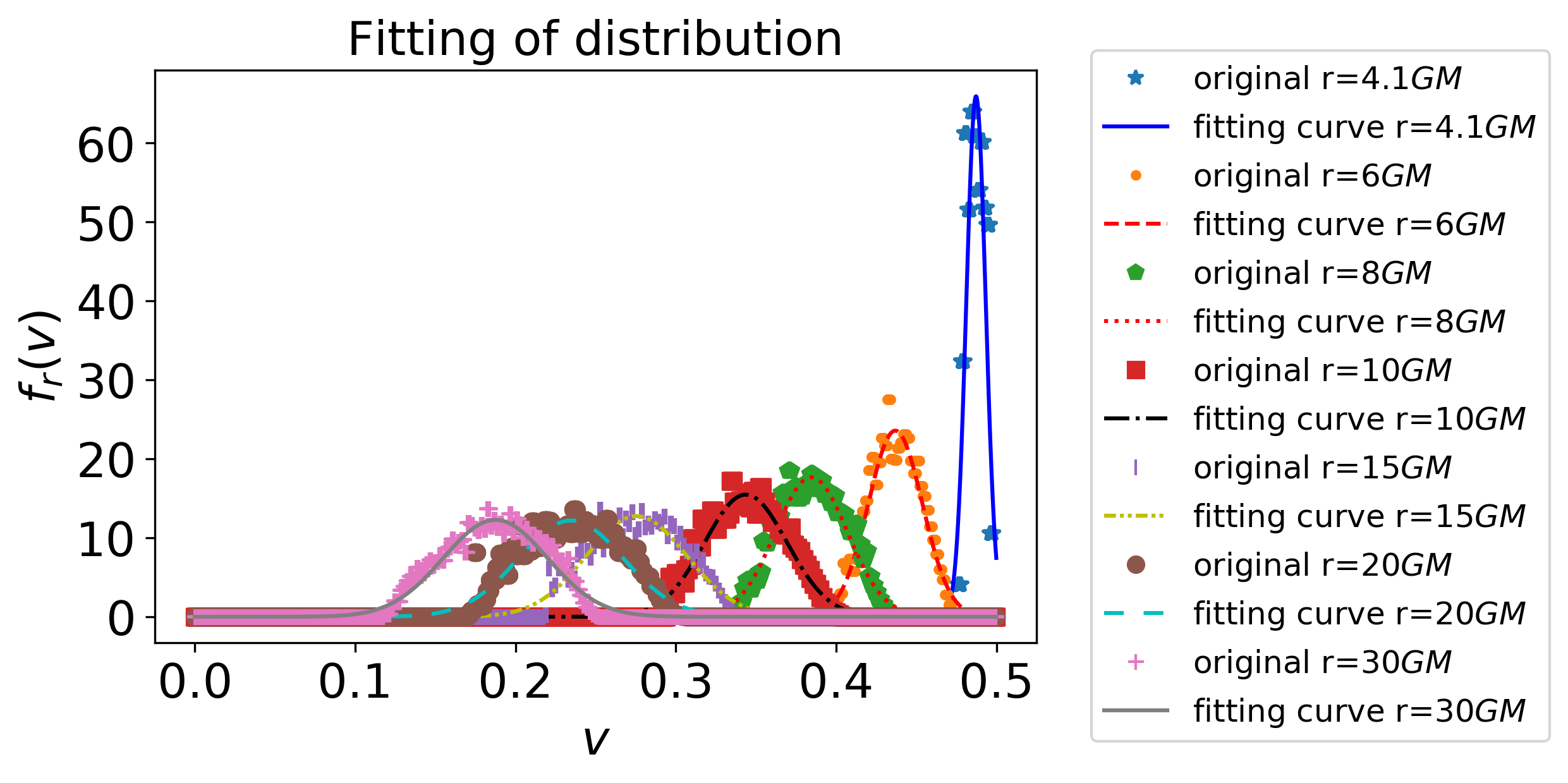}{(a)}
\includegraphics[width=3in]{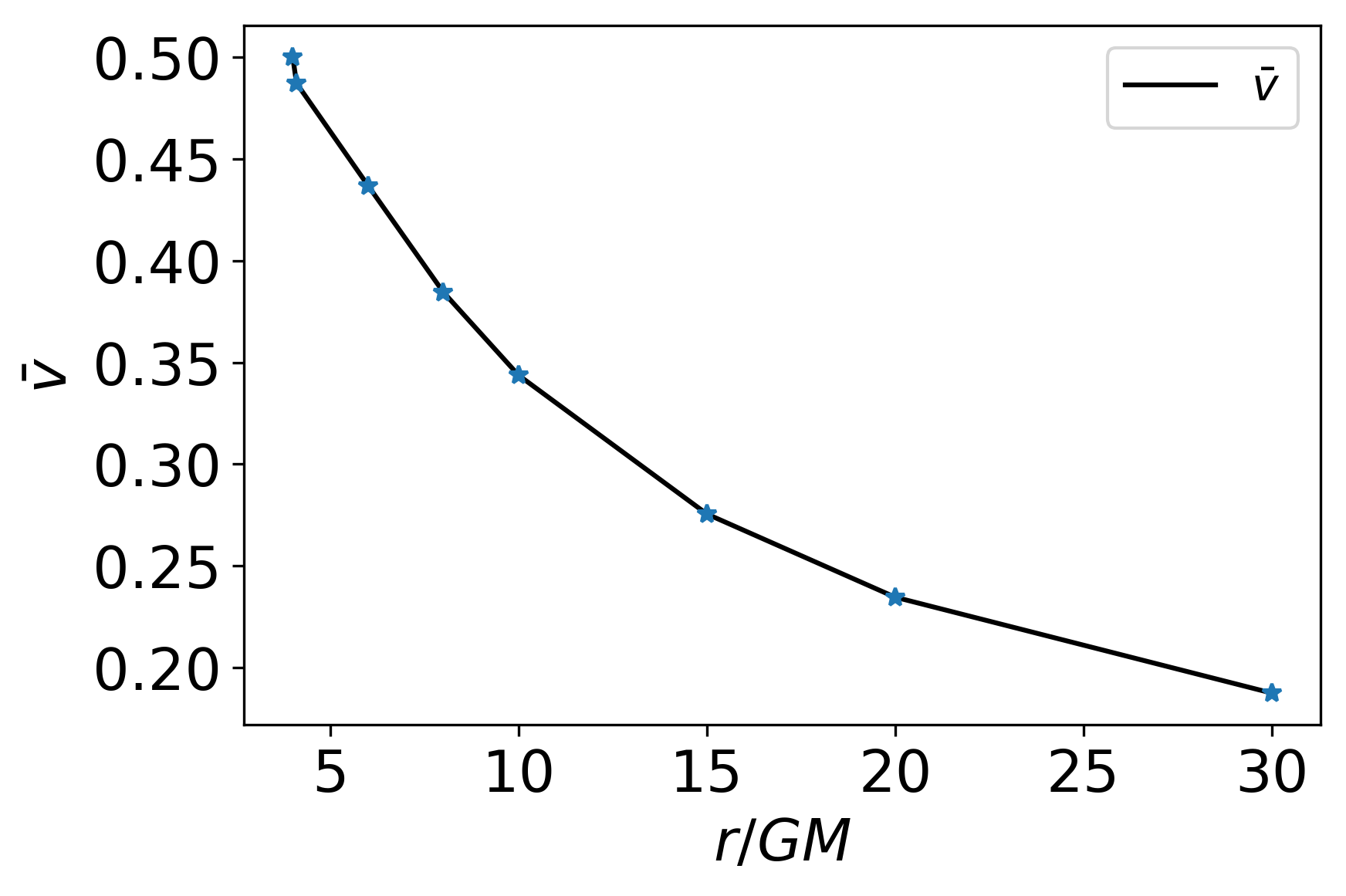}{(b)}
\includegraphics[width=3in]{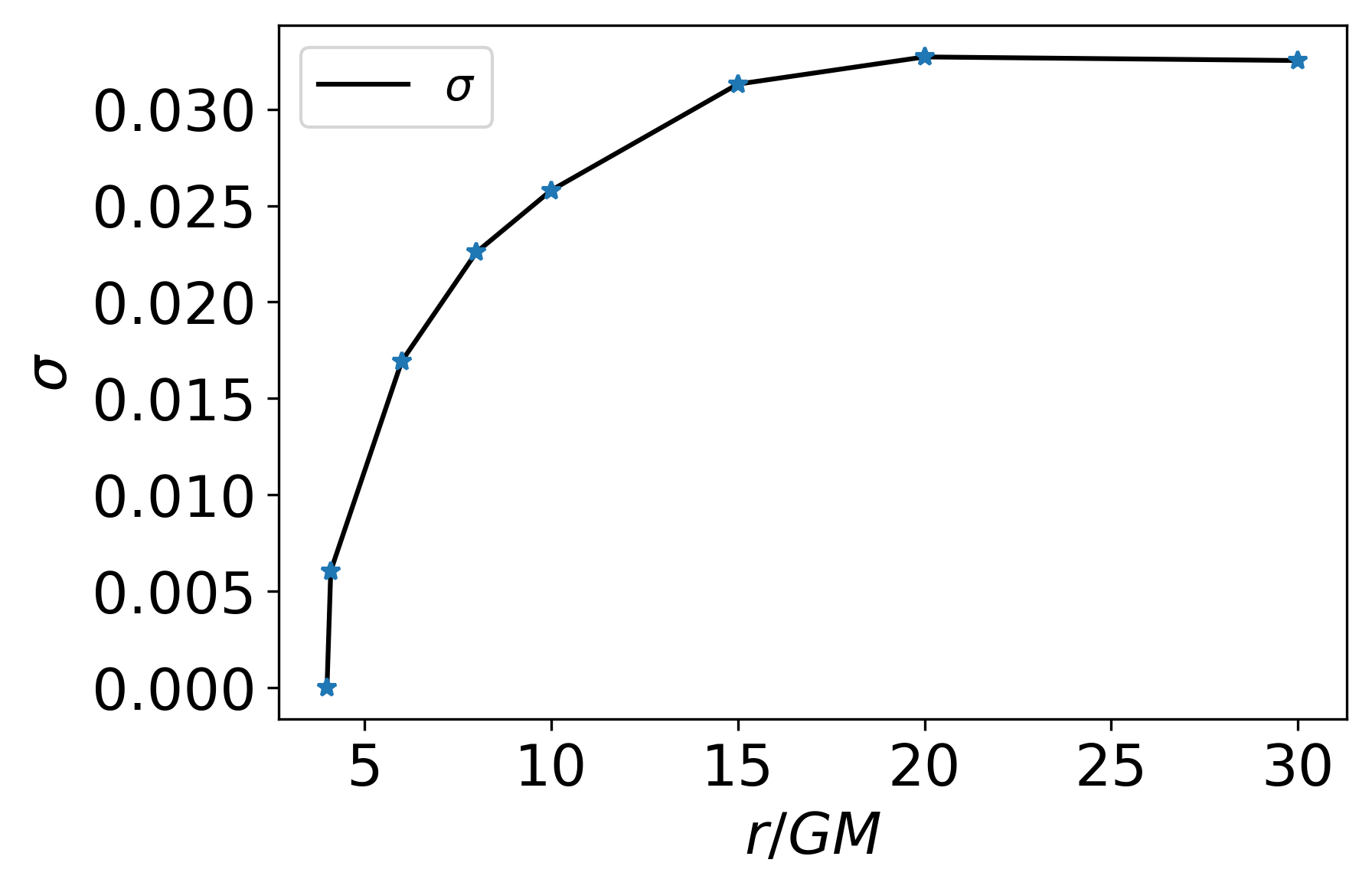}{(c)}
\caption{\label{fig:7} 
Fitting results from DM distribution around Galactic Center black hole. (a) Adapted distribution $f_r(v)$ at various distances with original scattering points. (b) The expected value $\bar{v}$ varies with $r$. (c) The variance $\sigma$ varies with $r$.}
\end{figure*}

In Figs.~(\ref{fig:1}-\ref{fig:3}) we show the various density distributions over DM particles' velocities and/or positions from center for three cases. In each figure, the four plots correspond to density function over velocity at different fixed positions, density function over position with upper bounds on the velocity, velocity dispersion at different positions and two-dimensional density distribution over position and velocity, respectively. Besides the overall factor, the common features displayed in all three cases suggest the scale invariance of density distribution. Thus, we only elaborate the first case in Fig.~\ref{fig:1}. 

In plot (a), we show how the density changes as we put an upper bound on the DM's velocity at three locations, $r=10GM,20GM$ and $30GM$. It is seen that, as the velocity approaches $v_{max}$, the density reaches its maximum. It also indicates that at location closer to the black hole DM particles are more focused on larger velocities. This can be more easily understood from the velocity distributions in plot (c) which exhibits that both central values and widths of the velocity distributions are different at three locations. In plot (b), we demonstrate the density function over distance, only summing particles up to some velocities, $v/v_{max}$. These curves show that DM density increases as approaching black holes but goes to zero at $r=4GM$ as no stable orbit exists $r<4GM$, and the maximal density is around $r\simeq 8GM$. Plot (d) is a two-dimensional density distribution over position and the velocity of DM up to some value, which is useful in the evaluation of DF of DM spike. 

Comparing Fig.~\ref{fig:1} with Fig.~\ref{fig:2}, we find that the spike in the vicinity of a black hole with mass of $10^4 M_\odot$ is about three orders of magnitude higher than galactic central black hole at the same normalized distance $\bar{r}= r/GM$. The difference in Fig.~\ref{fig:2} and Fig.~\ref{fig:3} shows that the high densities are closely related to the initial DM halo profile. The power-law halo can generate a spike whose density is about two orders of magnitude higher than that from Hernquist halo, although both give similar mass for the galactic dark halo. 

{Our results are different from Michel accretion under spherical symmetry~\cite{10.1111/j.1365-2966.2011.19258.x, 10.1111/j.1365-2966.2011.18687.x}. It is triggered by the different considerations of physical conditions. 
First, the premise of adiabatic growth requires that the orbital evolution of DM particles is on a timescale short compared to that of accretion from black hole. During the adiabatic growth, the variation of black hole mass is tiny enough to be ignored. 
Therefore, the generation of spike leaded by adiabatic growth is 
theoretically earlier than Michel accretion. 
Secondly, the DM particles in spikes formed by adiabatic growth are in stable bound orbits, which guarantees the formation of spikes. In our results, the spike vanishes at $r=4GM$ in Schwarzschild case due to the constraints of phase space $(\varepsilon,L^2,L_z)$. 
In~\cite{10.1111/j.1365-2966.2011.19258.x, 10.1111/j.1365-2966.2011.18687.x}, the density field of DM is narrowed to $r=GM$, and those DM particles inside the black hole’s horizon are taken into consideration. 
This also leads to differences in the distribution of DM.
Finally, we study the velocity distribution of DM particles at various distances in relativistic spikes. In phase space at a certain distance $r$, many allowable orbital parameters can be obtained. Thus, the location will be passed by various orbital DM particles contributing to the density $\rho(r)$. Then we are able to examine the velocity distribution of DM at different distances. However, DM evolved over Michel accretion has been in a steady state, in which the velocity field is different from that in an adiabatic spike.}

For quantitative description and potential use, we also fit the density total profiles below the maximum velocity $\alpha = v/v_{max} = 1$ to this form: 
\begin{equation}\label{eqs35}
\rho(r,\alpha=1) = \frac{\kappa}{(r/GM)^\omega}\left(1-\frac{4GM}{r}\right)^\eta.
\end{equation}
This corresponds to the black curve in plot (b). Here $\kappa$ reflects the overall magnitude of density, $\omega$ represents the power-law change with distance, 
and $\eta$ controls the size of the spike approaching to black hole. Again, the profile vanishes as $r\le4GM$ due to the nonexistence of stable orbits of DM particles. The best fitted values of $\kappa$, $\omega$ and $\eta$ are listed in the last column in Tab.~\ref{tab:parameter}. We can see in all cases $\eta\simeq 2$ and $\omega\simeq 2$.

{Intrigued by the shape of velocity distribution in plot (c) in each figure, and combining the analysis of Eq.~\ref{eqs2} in Sec.~\ref{subsec:vd}, we fit it with a Gaussian distribution in which the parameters vary with $r$,}
\begin{equation}\label{eqs36}
f_r(v) = \frac{1}{\sqrt{2\pi}\sigma(r)}\exp\left(-\frac{(v-\bar{v}(r))^2}{2\sigma^2(r)}\right),
\end{equation}
where $f_r(v)$ denotes the velocity distribution of DM particles at distance $r$, $\bar{v}(r)$ is the mean velocity at $r$, and variance $\sigma(r)$ shows the scale of velocity dispersion. We show the fitting results of Galactic Center black hole in Fig.~\ref{fig:7}. 
Plot (a) shows the numerically calculated points and the fitting curve at several distances. 
{As shown in Fig.~\ref{fig:7}, the parameters in Eq.~\ref{eqs36} are only regulated by $r$ and Gaussian distribution matches well. Therefore, we can conclude that the velocity distribution of DM in Schwarzschild spacetime depends on $r$ solely.}
We find that the particle velocity approaches a single value 0.5 when $r \to 4GM$. This is because that the closest bound circular orbit is uniquely determined at $r=4GM$. The constants of motion $(\varepsilon,L)$ in phase space are $(1,4GM)$~\cite{PhysRevD.88.063522}. Substituting it into geodesic equations we can derive that the particle at $r=4GM$ has a circular motion with a velocity $v =0.5$~\cite{hobson_efstathiou_lasenby_2006}. The variations of $\bar{v}$ and $\sigma$ over distance $r$ are demonstrated in plot (b) and (c) in Fig.~\ref{fig:7}, respectively. From the above analysis, we naturally set $\bar{v} = 0.5$ and $\sigma = 0$ at $r=4GM$.
Using such a distribution, we can simplify the calculation of DM spike density with velocity distribution close to Schwarzschild black hole by 
\begin{equation}\label{eqs37}
\rho(r, < v)=\rho(r, < 0.5)\int_{0}^{v} 4\pi v'^2 f_r(v')dv'.
\end{equation}

\begin{figure*}[ht!]
\includegraphics[width=5in]{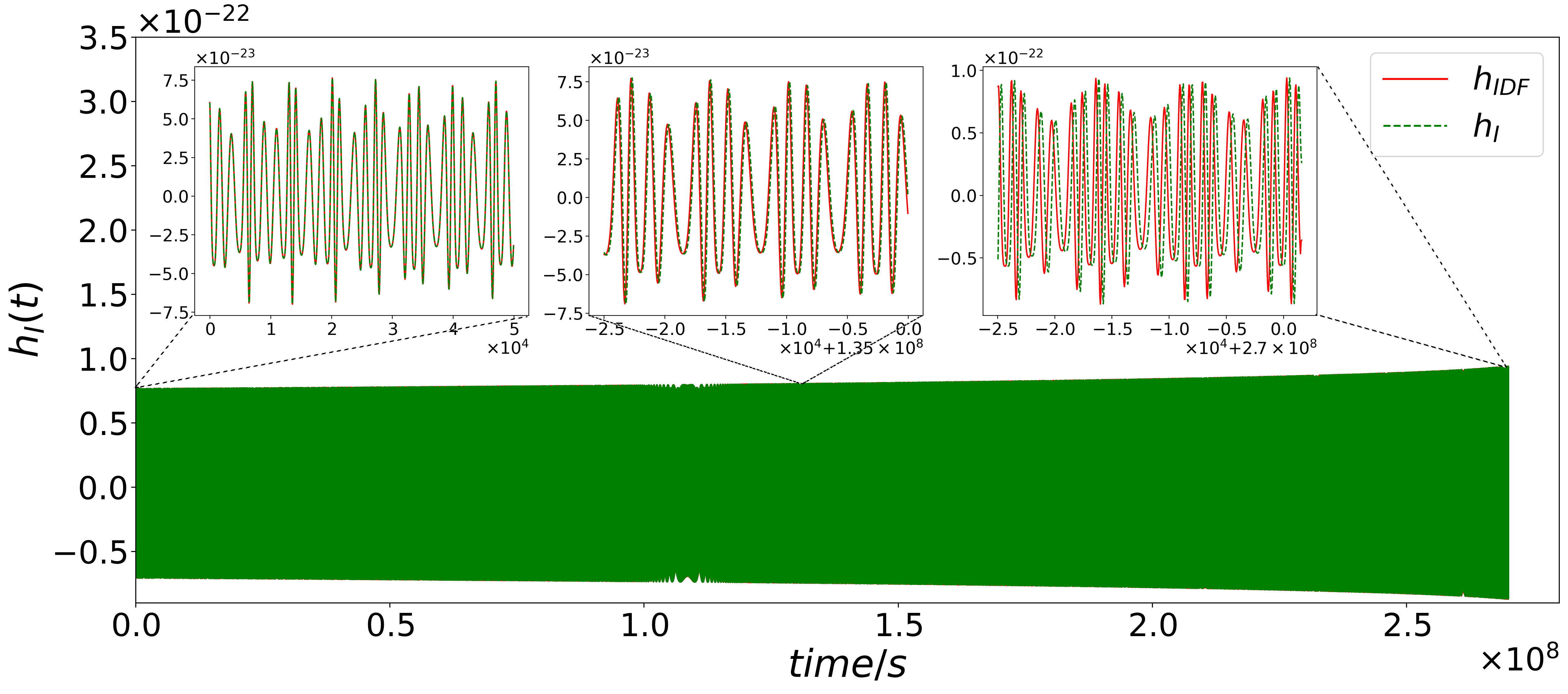}
\includegraphics[width=5in]{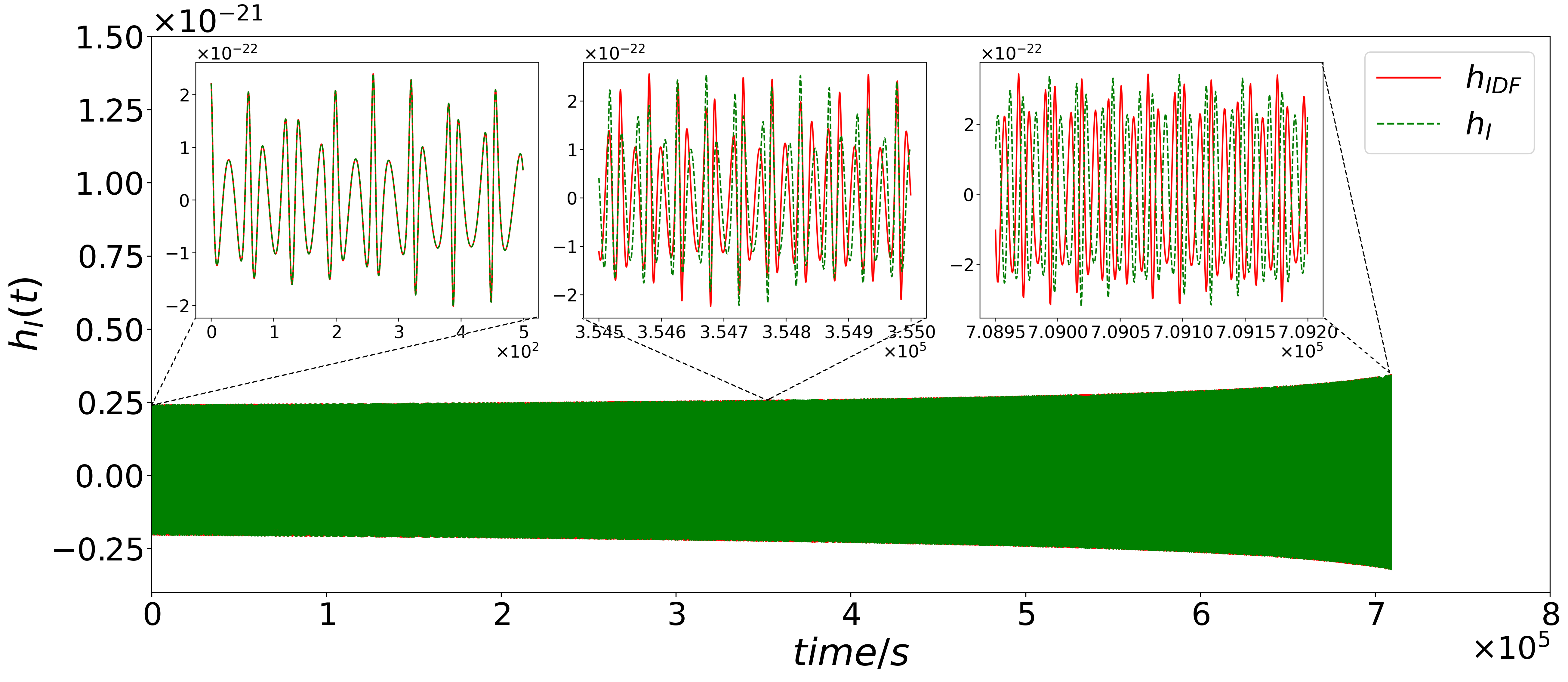}
\includegraphics[width=5in]{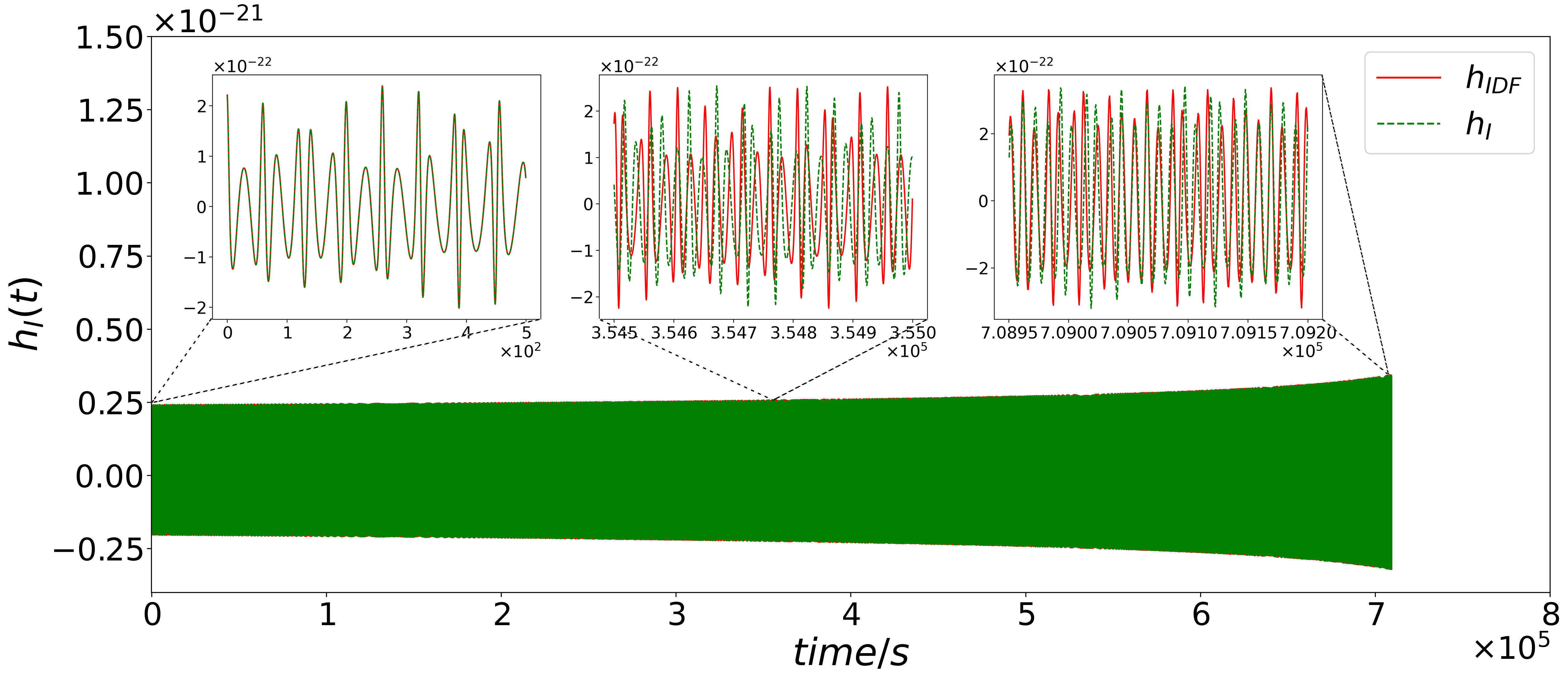}
\caption{\label{fig:5} GW waveform $h_{I}$ of EMRI with: (Upper) $(m, M, e_0, p_0, d) = (10M_\odot, 4.6\times10^6M_\odot, 0.2 ,8.5GM, 1~$Gpc) in a spike with an initial Hernquist dark halo, $(M_{halo},r_s) = (10^{12}M_\odot,20~$kpc). (Middle) $(m,M,e_0,p_0,d)=(10M_\odot, 10^4M_\odot, 0.3 ,30GM, 0.1~$Gpc) in a spike with an initial Hernquist dark halo, $(M_{halo},r_s) = (4.5\times10^8M_\odot,1.85~$kpc). (Lower) $(m,M,e_0,p_0,d)=(10M_\odot, 10^4M_\odot, 0.3 ,30GM, 0.1~$Gpc) in a spike with an initial power-law dark halo,  $(M_{halo}, r_s, \gamma) = (7.3\times10^8M_\odot,1.85~$kpc, 7/4). The solid red lines (dotted green) represent GW waveform with (without) DF.}
\end{figure*}

Finally we are in a position to discuss the effects of DF from DM spikes on GWs. We illustrate with two typical EMRI systems with physical parameters $(m, M, e_0, p_0, d)=(10M_\odot, 4.6\times10^6M_\odot, 0.2, 8.5GM, 1$Gpc) and $(10M_\odot, 10^4M_\odot, 0.3, 30GM, 0.1$Gpc), respectively. We show the time-domain waveforms of $h_{I}$ in Fig.~\ref{fig:5} for three cases. The solid red lines (dotted green) represent GW waveform with (without) DF. The signals terminate when the orbiting compact objects nearly plunge (the instantaneous orbits become unstable). The duration is about 8 years in the upper panel of Fig.~\ref{fig:5},  and about one week in the other two cases.

We can observe in Fig.~\ref{fig:5} that there is visible difference between GW waveforms with/without DF of DM spike in each case. The difference accumulates with time and in all three cases become sizable at plunge, which has been shown in the three small plots in each panel. We can also see that the effects of DF are stronger for smaller black holes. This can be understood as the DM density is larger near smaller black holes, as discussed above. Note that generally the difference also depends on the physical parameters of EMRIs. Here we just demonstrate with several examples. For more quantitative parameter estimations, more dedicated analysis will be needed, which is beyond the scope here and will be pursued in future work.

\section{SUMMARY} \label{sec:CON}

We have evaluated the velocity and density distributions of DM particles in spikes around Schwarzschild black holes in the relativistic case.  
By correlating the dark halo parameters with the mass of the central black hole, we have calculated the density of a DM spike from adiabatic growth of a black hole within a specific initial dark halo, such as Hernquist profile and power-law profile. We have found that the density profile of the DM spike is scale invariant in its shape, but larger in magnitude at the same normalized distance $\bar{r}= r/GM$ for smaller central black hole within the same initial dark halo. In Hernquist dark halo, the DM spike surrounding a black hole with mass of $10^4 M_\odot$ is three orders of magnitude denser than that around the supermassive black hole in Milky Way. 
Moreover, the densities of spikes for different initial conditions can vary significantly. Near the black hole with mass of $10^4 M_\odot$, the density in a spike that grows from a power-law halo is about two orders of magnitude higher than that from a Hernquist halo.

We have also visualized the velocities of DM particles in the spike and fitted the dispersion with Gaussian distribution. This has simplified the subsequent DF calculation, where relativistic effect is taken into account. The result shows that the velocities of DM particles closer to the black hole tend to be narrowly distributed, while the velocities of particles far away scatter more widely. 

Finally we have investigated the GW waveform of a stellar-mass compact object orbiting a massive black hole within a DM spike along general elliptical orbit.  
Those DM particles that are slower than the orbiting objects generate gravitational drag that can change the evolution of the object. Our results show such a dynamical friction effect causes visible phase change in the waveforms of GWs emitted from EMRIs, which is one of main sources for future GW detectors in space, such as LISA and Taiji.

\begin{acknowledgments}

This work is supported by the National Key Research and Development Program of China (Grant No.2021YFC2201901), the National Natural Science Foundation of China (Grant No.12347103) and the Fundamental Research Funds for the Central Universities. 
We acknowledge the use of \texttt{NumPy} \cite{Harris_2020}, \texttt{SciPy} \cite{Virtanen_2020}, and \texttt{Matplotlib} \cite{4160265} for numerical calculations and data visualization.
\end{acknowledgments}
\bibliography{citation}
\bibliographystyle{apsrev}

\end{document}